\documentclass[journal,10pt]{IEEEtran}
\pdfoutput=1
\IEEEoverridecommandlockouts
\usepackage{amsbsy}
\usepackage{amsthm}
\usepackage{amssymb}
\usepackage{amsmath}
\usepackage{multirow}
\usepackage{epsfig}
\usepackage{subfigure}
\usepackage{graphics}
\usepackage{multicol}
\usepackage{pifont}
\usepackage{makecell}
\usepackage{epstopdf}
\usepackage{balance}
\usepackage{xcolor}
\usepackage{cite}
\usepackage{color}
\usepackage{algorithm}
\usepackage{algorithmic}
\usepackage{textcomp}
\usepackage{subfigure}
\usepackage{subfloat}
\usepackage{booktabs}
\usepackage{bm}
\newcommand{\cmark}{\ding{51}}%
\begin{document}

\title{Joint Beamforming Design for Dual-Functional MIMO Radar and Communication Systems Guaranteeing Physical Layer Security}

\author{Fuwang~Dong,
        Wei~Wang,~\IEEEmembership{Senior~Member,~IEEE}, Xin~Li, Fan Liu, Sheng~Chen,~\IEEEmembership{Fellow,~IEEE}, and Lajos Hanzo,~\IEEEmembership{Life Fellow,~IEEE}
\thanks{This work is supported in part by the National Natural Science Foundation udner Grant  62271163, in part by the Fundamental Research Funds for the Central Universities (3072022QBZ0401, 3072021CFT0404). F. Liu would like to acknowledge the financial support of the National Natural Science Foundation of China under Grant 62101234, as well as of the Young Elite Scientist Sponsorship Program by the China Association for Science and Technology (CAST) under Grant No. YESS20210055. L. Hanzo would like to acknowledge the financial support of the Engineering and Physical Sciences Research Council projects EP/W016605/1 and EP/P003990/1 (COALESCE) as well as of the European Research Council's Advanced Fellow Grant QuantCom (Grant No. 789028). (\textit{Corresponding author: Wei Wang.})}
\thanks{Fuwang Dong, and Fan Liu are with the Department of Electronic and Electrical Engineering, Southern University of Science and Technology, Shenzhen 518055, China (email: dongfw@sustech.edu.cn; liuf6@sustech.edu.cn)
	
Wei Wang, and Xin Li are with the College of Intelligent System Science and Engineering, Harbin Engineering University, Harbin, 150001, China (email: wangwei407@hrbeu.edu.cn; xinxin\_forever@126.com ).

Sheng Chen, and Lajos Hanzo are with the School of Electronic and Computer Science, University of Southampton, Southampton SO17 1BJ, U.K. (email: sqc@ecs.soton.ac.uk; lh@ecs.soton.ac.uk).}}
\maketitle

\begin{abstract}
The dual-functional radar and communication (DFRC) technique constitutes a promising next-generation wireless solution, due to its benefits in terms of power consumption, physical hardware, and spectrum exploitation. In this paper, we propose sophisticated beamforming designs for multi-user DFRC systems by additionally taking the physical layer security (PLS) into account. We show that appropriately designed radar waveforms can also act as the traditional artificial noise conceived for drowning out the eavesdropping channel and for attaining increased design degrees of freedom (DoF). The joint beamforming design is formulated as a non-convex optimization problem for striking a compelling trade-off amongst the conflicting design objectives of radar transmit beampattern, communication quality of service (QoS), and the PLS level. Then, we propose a semidefinite relaxation (SDR)-based algorithm and a reduced-complexity version to tackle the non-convexity, where the globally optimal solutions are found. Moreover, a robust beamforming method is also developed for considering realistic imperfect channel state information (CSI) knowledge. Finally, simulation results are provided for corroborating our theoretical results and show the proposed methods' superiority.      
\end{abstract}


\IEEEpeerreviewmaketitle

\section{Introduction}\label{introduction}

The proliferation of wireless mobile services exhibits an exponential trend, leading to a scarcity of spectral resources and to escalating spectrum prices. For example, it has been reported that the number of connected devices is expected to be 80 billion by 2030 with an annual growth rate of around 25\%, and that of the active Internet of Things (IoT) devices will reach 24.1 billion by 2030 \cite{Online1,LiuFan2020}. Recently, the concept and scope of \textit{Integrated Sensing and Communication (ISAC)} technology have been formally defined in \cite{LiuISAC2021,Cui2021}, enabling sensing and communication simultaneously in the same frequency band or/and hardware platform, which can significantly improve the resource utilization. Due to the numerous advantages offered by ISAC, it is envisioned to be a promising technique in terms of supporting autonomous vehicles \cite{Ma2020,PK2018} and the IoT in 6G wireless networks \cite{IoT2020}.

There are two main ISAC categories in terms of transmitted signal: radar and communication spectrum coexistence and dual functional radar-communication (DFRC) \cite{NCL2021,LeZheng2019}. In this paper, we consider a DFRC system, which transmits dual-functional signals/waveforms from a single hardware platform, to gain benefits from joint sensing and signaling operations via real-time cooperation. The main motivation of transmit beamforming is to synthesize multiple beams towards both the communication users and the radar targets by exploiting the associated spatial degrees of freedom (DoF). In \cite{LiuFan2018}, the authors considered the radar targets as virtual downlink users encountering a line of sight (LoS) channel. Therefore, the beamforming matrix was designed for closely matching the desired radar beampattern, while simultaneously guaranteeing the signal to interference and noise ratio (SINR) attained by the downlink users. Furthermore, the authors of \cite{LiuFanDual2018,Hezishu2020} studied the associated symbol/waveform level probing signal design issues, where the multi-user interference energy was minimized under the similarity and constant modulus constraints of the radar waveform. However, the above-mentioned schemes only utilize the communication waveform as the DFRC waveform to implement target detection, hence leading to a DoF reduction, thereby to a radar performance degradation. To this end, the authors of \cite{LiuXiang2020} firstly proposed a jointly precoded individual communication and radar waveforms based scheme, where the communication signal can be regarded as a special case relying on nullifying the dedicated radar waveforms. Therefore, by exploiting the inherent advantages of the radar waveform, the DoF erosion can be efficiently compensated, hence resulting in target detection performance improvements, especially for a small number of downlink users.  

\begin{table*}[!htbp]
	\centering
	\caption{Our contributions in contrast to the state-of-the-art.}
	\label{table1}
	\setlength{\tabcolsep}{2mm}{
		\begin{tabular}{|l|c|c|c|c|c|c|} 
			\Xhline{0.75pt}
			&\cite{LiuFan2018}  & \cite{LiuXiang2020} & \cite{JC2018} & \cite{DSP2018} & \cite{NanSu2021} & Our work \\
			\Xhline{0.75pt}
			Secure Transmission &   &  & \cmark & \cmark & \cmark &  \cmark \\ 
			\Xhline{0.75pt}
			Jointly precoded communication and radar waveforms &   & \cmark &  &  &  & \cmark \\ 
			\Xhline{0.75pt}
			Precoder design rather than covariance matrix & \cmark &  \cmark & \cmark &  &  & \cmark \\ 
			\Xhline{0.75pt}
			Radar beampattern optimization & \cmark & \cmark &  &  &  &  \cmark \\			
			\Xhline{0.75pt}
			Multiple users & \cmark &  \cmark &  &  & \cmark & \cmark \\ 
			\Xhline{0.75pt}
			Imperfect CSI estimations  &   &  &  &  & \cmark & \cmark \\			
			\Xhline{0.75pt}
			Multiple eavesdroppers &  &  &  &  &  & \cmark \\ 
			\Xhline{0.75pt}
			Using radar signal as artificial noise &  &  &  &  &  & \cmark \\ 
			\Xhline{0.75pt}
			Tight solution for PLS design &  &  &  &  &  & \cmark \\ 
			\Xhline{0.75pt}
	\end{tabular}}
\end{table*} 

Another critical problem in the DFRC system, which has been largely overlooked in the relevant literature, is how to guarantee the privacy and security of the desired information~\cite{NW2019}. The DFRC base station (BS) transmits the dual-functional probing waveform for detection purposes, but also sends confidential information to the targets. Evidently, private information might be leaked to the targets, which may act as potential eavesdroppers (Eves). Recently, several schemes have been proposed for guaranteeing secure data transmission by exploiting constructive interference \cite{NS2022}, frequency hopping \cite{KW2022}, and additional artificial noise (AN) \cite{JC2018,NanSu2021,DSP2018}, etc. As a low complexity yet powerful technique, the AN method has been widely harnessed in the communication community for enhancing the physical layer security (PLS). The basic principle of AN-aided secure transmission is that of contaminating the transmit signal by well-designed AN to degrade Eve's reception without affecting the legitimate users (LUs) \cite{SG2008}.

In \cite{JC2018}, several optimization problems, including secrecy rate maximization, target return SINR maximization, and transmit power minimization were formulated for a DFRC system in the presence of a single target and a single communication receiver. To tackle the non-convexity of the secrecy rate expression, an approximate algorithm based on the first-order Taylor expansion was proposed, which however resulted in a performance gap between the original non-convex problem and the approximated one. The authors of \cite{DSP2018} considered a unified joint passive radar and communication system, where the SNR at the passive radar receiver was maximized, while keeping the secrecy rate above a certain target. Moreover, several practical constraints, such as realistic target direction estimation and imperfect channel state information (CSI) were taken into account in the associated robust beamforming proposal of \cite{NanSu2021}. However, at the time of writing, most of the contributions on secure DFRC systems have the following two drawbacks: (1) They only design the covariance matrix of the AN, yet no further analysis of the DFRC system's radar detection is offered; (2) Several relaxation algorithms are used such as Taylor expansions or semidefinite relaxation (SDR) techniques, but the performance loss compared to the original non-convex problem is overlooked. 

Motivated by filling the above-mentioned knowledge gap in the literature, we develop jointly precoded communication and radar waveforms for secure transmission in a multiple-input multiple-output (MIMO) DFRC system inspired by \cite{LiuXiang2020}, serving multiple LUs and detecting the targets simultaneously. On one hand, the DFRC platform relying on the ISAC technique eliminates duplication in the system's hardware. On the other hand, the bespoke transmit signals can simultaneously meet the requirements of radar, communications, and PLS, circumventing redundancy in the resource consumption for each functionality, hence also the power dissipation. Compared to the current DFRC schemes such as those in \cite{LiuFan2018,NanSu2021,LiuFanDual2018,Hezishu2020}, our method achieves superior radar detection performance thanks to the increased DoFs attained by the additional radar waveforms. In contrast to \cite{LiuXiang2020}, the PLS level is also considered in our work, where the targets may act as potential Eves. The radar waveforms conveying no confidential information may also be exploited as the AN imposed on the communication signals for contaminating the eavesdropping channels. The main contributions of this paper are summarized as follows, and they are also boldly and explicitly contrasted to the literature at a glance in Table 1.

\begin{itemize}
\item We develop jointly precoded communication and radar waveforms for secure transmission. Specifically, the AN of traditional PLS designs can be replaced by bespoke radar signals specifically designed for inflicting interference upon the Eves, whilst additionally increasing the DoF available for target detection.
   
\item We formulate the joint beamforming design as a non-convex optimization problem under the consideration of both radar, communication and security performance. An SDR-based and the associated low complexity algorithms are also conceived for tackling the non-convexity of the problem, where we prove that the relaxation used in our scheme is tight.

\item We propose a robust beamforming design for the more practical scenarios of imperfect estimations, including the uncertain target directions and the imperfect CSI acquired for the LUs. We also show that the globally optimal reconstruction method proposed for ideal scenarios still applicable to our robust beamforming scheme.

\item We analyze the performance trade-offs among radar, communication and PLS both theoretically and by simulation for providing new insights into flexible beamforming. 
   
\end{itemize}

The rest of this paper is organized as follows. In Section \ref{SMPF}, we establish the mathematical model of joint communication and radar signal transmission and introduce the performance metrics of radar detection, multiuser communication, and system security, respectively. The proposed SDR-based beamforming and the low complexity ZF-based algorithms are characterized in Section \ref{JBCR}. Furthermore, Section \ref{robust} provides our robust beamforming method relying on imperfect CSI knowledge, while the performance vs. the complexity of the proposed algorithms is analyzed in Section \ref{USS}. Finally, our simulation results and conclusions are provided in Section \ref{Simulation} and \ref{Conclusion}, respectively.  
\begin{table}[!t]
	\centering
	\caption{Frequently Used Symbols}
	\label{table2}
		\setlength{\tabcolsep}{2mm}{
			\begin{tabular}{ll} 
			\toprule
			Notation & Description\\
			\midrule
			$\textbf{R}$   & Covariance matrix of the transmitted signals \\
			$\textbf{H}$   & Communication CSI matrix \\
			$\textbf{W}_r$ ($\textbf{W}_c$)  & Radar (Communication) beamforming matrix \\
			$\Gamma_e$ ($\Gamma_c$) & SINR threshold at Eves (LUs) \\
			$K$ & Number of the LUs \\
			$Q$ & Number of the targets (Eves)  \\
			$M$ & Number of antennas  \\
			$\beta$ & Path-loss coefficient for radar channel  \\
			$L_r(\textbf{R},\alpha)$ & Least square function for MIMO radar beampattern\\
			$\gamma_k$ ($\tilde{\gamma}_q$) & SINR of the $k$-th LU (the $q$-th Eve)  \\
			\bottomrule
	        \end{tabular}}
	\end{table}

The notations used in this paper are as follows. Upper-case $\textbf{A}$ (lower-case $\textbf{a}$) bold characters denote matrices (column vectors), and lower case normal letters $a$ are scalars; $(\cdot)^T$, $(\cdot)^*$ and $(\cdot)^H$ represent the transpose, conjugate and complex conjugate transpose operations respectively; $|a|$ and $\|\textbf{a}\|_2$ stand for the magnitude of a scalar $a$ and the $\ell_2$-norm of the vector $\textbf{a}$;  $\mathbb{E}\{\cdot \}$ is the statistical expectation; $\textrm{diag}\{\textbf{a}\}$ stands for a diagonal matrix using the elements of $\textbf{a}$ as its diagonal elements; for a matrix $\textbf{A}$, $[\textbf{A}]_{[i,j]}$ denotes the $(i,j)$th element; $\textbf{A}_{[:,1:k]}$ and $\textbf{A}_{[1:k,:]}$ represent the sub-matrices containing the first $k$ columns and rows of $\textbf{A}$ respectively; $\textbf{I}_M$ is the $n$-dimensional identity matrix and $\textbf{0}_{M \times N}$ is the $M \times N$ matrix having all-zero entries. Frequently used symbols in this paper are summarized in Table \ref{table2}.  
 
\section{ System Model and Performance Metrics}\label{SMPF}
\subsection{Transmission and Reception Signal Model} 
\begin{figure}[t]
	\centering
	\includegraphics[width=3in]{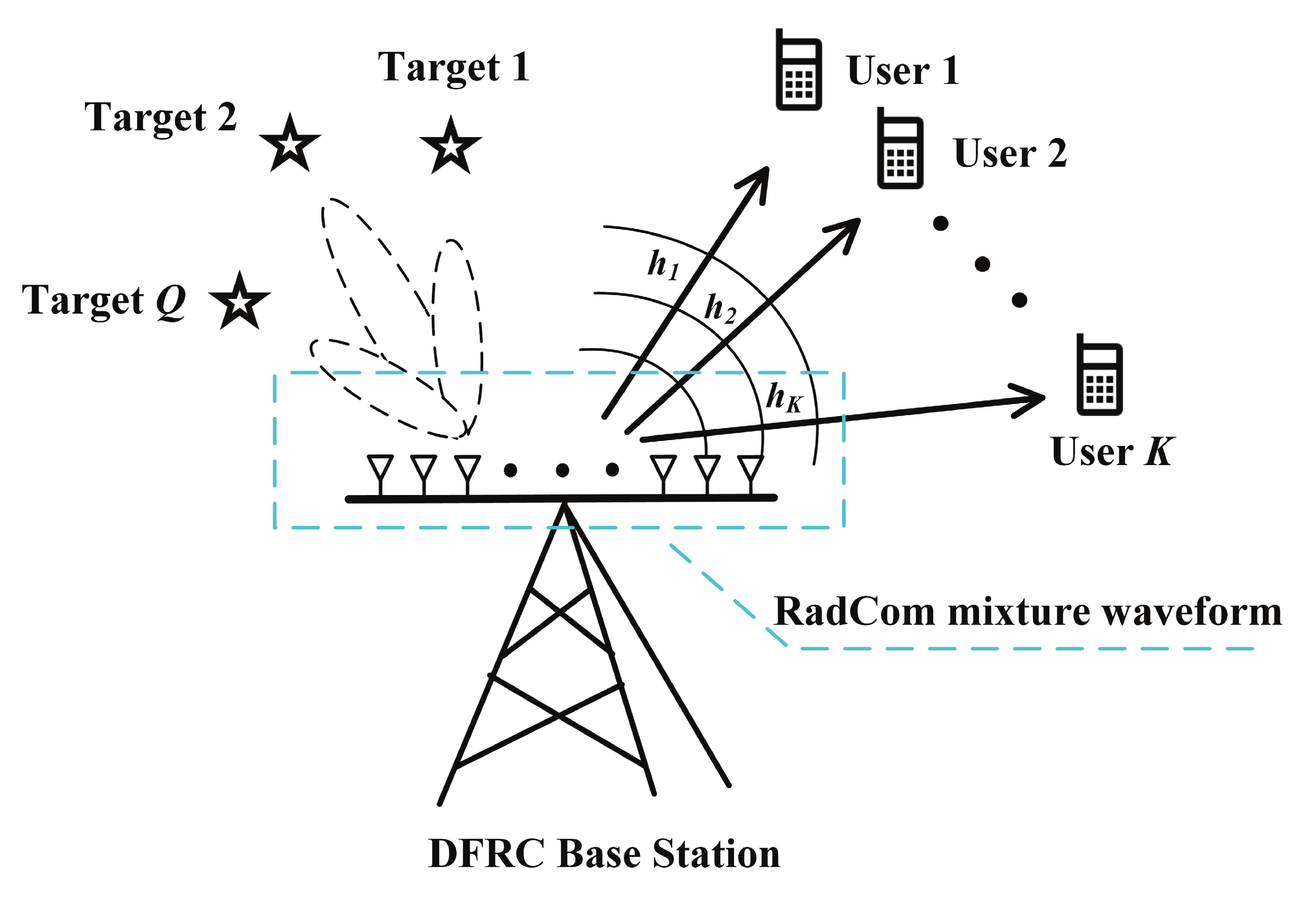}
	\caption{The DFRC system detects the targets (Eves) and serves downlink users by transmitting mixture waveform.}
	\label{system}
\end{figure}

As shown in Fig. \ref{system}, a colocated MIMO BS transmits DFRC signals to detect $Q$ targets and $K$ LUs simultaneously. For the consideration of our PLS design, all the targets considered are non-cooperative, such as unmanned aerial vehicle (UAV) which are regarded as the potential Eves at the same time. We assume that the BS is equipped with $M$ antennas arranged in a uniform linear array (ULA), and all the Eves and LUs have a single antenna. The proposed beamforming design can be readily extended to multi-antenna scenarios.
	
Following \cite{LiuXiang2020}, the discrete-time transmitted signal at time slot $n$, which is the weighted sum of the communication signals and radar waveforms, can be expressed as
\begin{equation}\label{signal}
\textbf{x}[n]=\textbf{W}_r\textbf{s}[n]+\textbf{W}_c\textbf{c}[n], \kern 2pt n=0, 1, \cdots, N-1, 
\end{equation}  
where $\textbf{s}[n]=[s_1[n],\cdots,s_M[n]]^T $ represents the individual radar signals and $\textbf{c}[n]=[c_1[n],\cdots,c_K[n]]^T$ stands for the $K$ parallel communication symbol streams intended for the LUs. $N$ is the total number of symbols. Furthermore, $\textbf{W}_r\in \mathbb{C}^{M \times M}$ and $\textbf{W}_c\in \mathbb{C}^{M \times K}$ denote the beamforming matrices (or precoders) designed for the radar waveforms and communication waveforms. The conventional transmit signal strategy which only exploits the communication signals for detection in \cite{LiuFan2018,NanSu2021,LiuFanDual2018,Hezishu2020}, can be regarded as the special case associated with $\textbf{W}_r=\textbf{0}$. In line with the literature, the following assumptions are stipulated for the transmitted signals (\ref{signal}).    
\begin{itemize}
\item Both the radar and communication signals have zero mean, and they are temporally white wide-sense stationary stochastic processes;
\item The radar and the communication waveforms are statistically independent, hence we have $\mathbb{E}\{\textbf{s}\textbf{c}^H \}=\textbf{0}_{M \times K}$; 
\item The $M$ radar waveforms are orthogonal to each other, then we have $\mathbb{E}\{\textbf{s}\textbf{s}^H \}=\textbf{I}_M$;            
\item The communication symbols transmitted to different LUs are uncorrelated, i.e., $\mathbb{E}\{\textbf{c}\textbf{c}^H \}=\textbf{I}_K$;    
\end{itemize}
Here, the signal power is normalized to unity. Thus, the covariance matrix of the transmitted signal can be written as
\begin{equation}\label{CMT}
\textbf{R}=\mathbb{E} \{ \textbf{x} \textbf{x}^H \} = \textbf{W}_r\textbf{W}_r^H+\textbf{W}_c\textbf{W}_c^H. 
\end{equation} 

Let $\textbf{y}=[y_1,y_2,\cdots,y_K]^T$ denote the received signal vector of all the LUs, which can be expressed by 
\begin{equation}\label{RecieveBob}
\textbf{y}=\textbf{H}\textbf{x}+\textbf{n}_c, 
\end{equation}
where $\textbf{H}=[\textbf{h}_1^*,\cdots,\textbf{h}_K^*]^T \in \mathbb{C}^{K \times M}$ is the channel matrix and $\textbf{h}_k$ represents the channel vector spanning from the BS to the $k$th LU, and $\textbf{n}_c\sim \mathcal{CN}(0,{\sigma_c ^2}\textbf{I}_K)$ denotes the additive white Gaussian noise (AWGN). Moreover, the targets of interest can be viewed as virtual downlink users located in the LoS channel of DFRC systems \cite{LiuFan2018}. Therefore, the signal received by the $q$th target (Eve) can be modeled as \cite{NanSu2021}
\begin{equation}\label{RecieveEve}
r_q=\beta_q\textbf{a}^H(\theta_q)\textbf{x}+n_e, 
\end{equation}  
where $\beta_q$ is the complex path-loss coefficient, $n_e$ is the AWGN with covariance $\sigma_e^2$, and $\textbf{a}(\theta)$ represents the ULA arrays' steering vector, which can be expressed as
\begin{equation} \label{steer}
\textbf{a}(\theta)=\frac{1}{\sqrt{M}} \left[1,e^{\jmath 2\pi\frac{d}{\lambda}\sin(\theta)},\cdots,e^{\jmath 2\pi(M-1)\frac{d}{\lambda}\sin(\theta)} \right]^T.
\end{equation} 
Here, $d$ is the antenna spacing, $\lambda$ is the carrier wavelength, and $\theta$ is the azimuth of the target.

The BS has to acquire the CSI for both LUs and Eves before the beamforming design. In general, the CSI marix $\textbf{H}$ of LUs can be obtained through channel estimation and feedback techniques \cite{Health2016}. By contrast, the CSI from the BS to the Eve is challenging to acquire, since the Eves tend to be passive in general. Fortunately, the sensing functionality of the DFRC signal can be exploited for estimating the azimuth and path-loss coefficient through radar parameter estimation techniques \cite{DoA1,DoA2}. Since we only focus on the beamforming design, the processes of radar parameter estimation and information demodulation are ignored in this paper. The elaborate details can be found in \cite{LiuFan2020,IEEE2011}. Before proceeding to our mathematical analysis, we have depicted in Fig. \ref{PLS_Flow} the flow of the analysis described in the sequel, which allows readers to grasp the overall structure of this paper at a glance. 

 \begin{figure}[!t]
	\centering
	\includegraphics[width=3in]{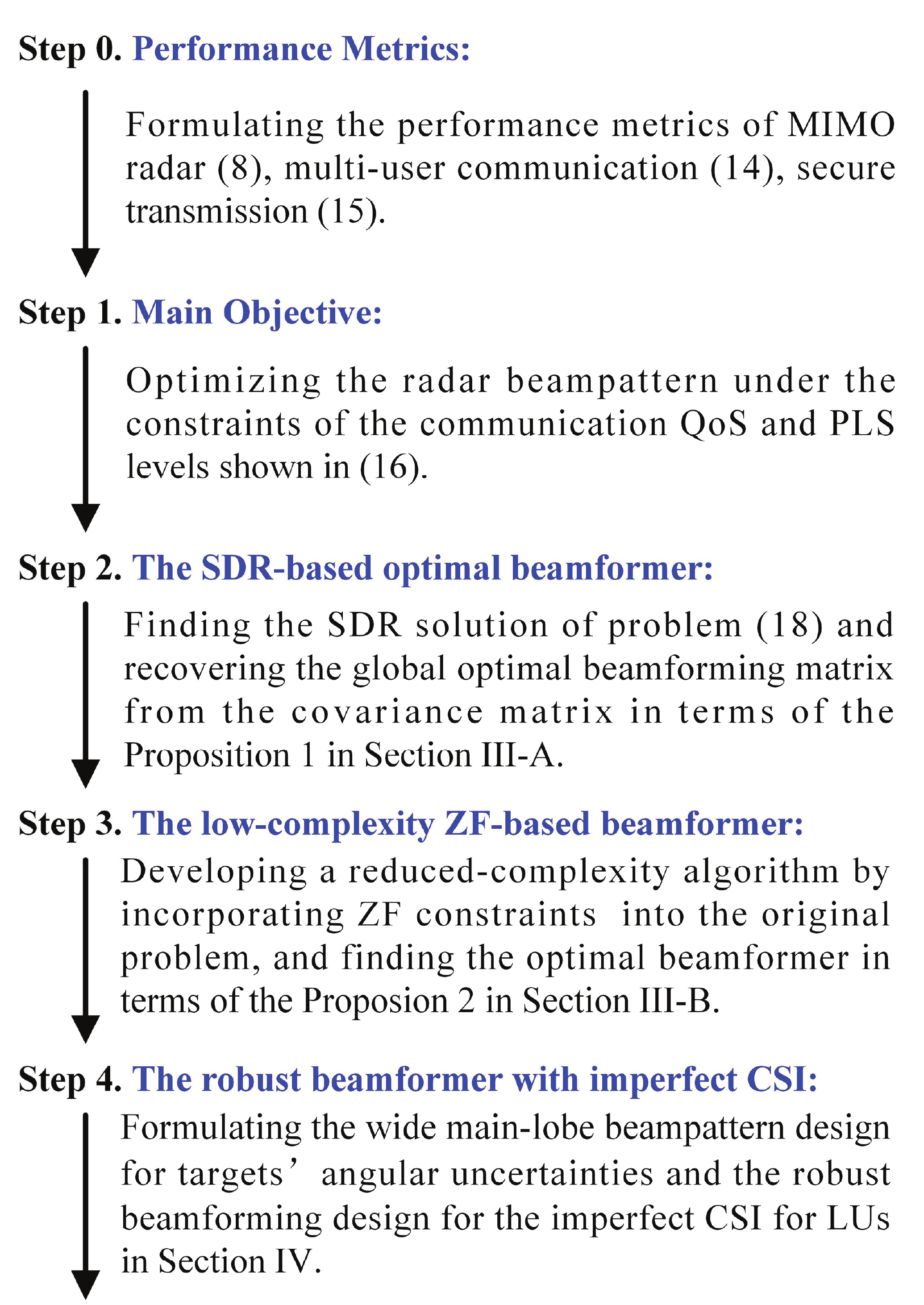}
	\caption{Flow of the mathematical analysis.}
	\label{PLS_Flow}
\end{figure}
  
\subsection{Performance Metrics}\label{PM}
In our proposed physical layer beamformer designed for secure transmission, some important properties related to the symbol-level waveform design \cite{Hezishu2020,LiuFanDual2018} are not considered, such as the radar's ambiguity function, the peak-to-average power ratio (PAPR), etc. Next, we introduce our performance metrics used for the target detection, for the communication quality of service (QoS), and for the PLS level, respectively.

(1) \textbf{\textit{Performance metric for MIMO radar:}} In general, there are two primary MIMO radar functions, namely detection and tracking. MIMO radar tends to create both spatially orthogonal waveforms and omni-directional beampatterns (i.e., $\textbf{R}=\textbf{I}$) for detecting the potential targets in the detection stage, since there is no prior information concerning the targets. Then, in the tracking stage, MIMO radar steers the beam to the target directions of interest acquired during the previous observations. Instead of maximizing the SINR at radar receiver \cite{ISJ3}, we focus on the radar transmit beampattern performance. The synthesized radar beampattern at azimuth $\theta$ can be formulated as   
     
\begin{equation} \label{beampattern}
P(\theta;\textbf{R})=\mathbb{E}\{\textbf{a}^H(\theta)\textbf{x}\textbf{x}^H\textbf{a}(\theta) \}=\textbf{a}^H(\theta)\textbf{R}\textbf{a}(\theta).
\end{equation} 
Additionally, the cross-correlation pattern between direction $\theta_1$ and $\theta_2$ can be written as
\begin{equation} \label{beampattern2}
P_c(\theta_1,\theta_2;\textbf{R})=\textbf{a}^H(\theta_2)\textbf{R}\textbf{a}(\theta_1).
\end{equation}  
The objectives of beamformer design for MIMO radar include the following \cite{PS2007}
\begin{itemize}
	\item Optimize the beampattern over the sectors of interest to concentrate the signal power while maintaining a low sidelobe level;
	\item Reduce the cross-correlation pattern over the set of target angles to achieve an excellent adaptive performance;
\end{itemize}
To this end, we adopt the loss function defined in terms of the least squares as our performance metric for MIMO radar, which is formulated as
\begin{equation} \label{RdarMetric}
L_r(\textbf{R},\alpha)=L_b(\textbf{R},\alpha)+\eta L_c(\textbf{R}),
\end{equation}      
where $\eta$ is the weighting factor representing the relative importance of the two terms based on the associated practical requirements. The first term represents the mean squared error between the designed and desired beampatterns, which can be formulated as
\begin{equation} \label{DDD}
L_b(\textbf{R},\alpha)=\frac{1}{L}\sum \limits_{l=1}^L | \alpha \Phi(\theta_l)-P(\theta_l;\textbf{R})|^2.
\end{equation}
Here, $\alpha$ is a scaling factor, $\Phi(\theta)$ denotes the desired transmit beampattern, and $\{\theta_l\}_{l=1}^L$ represents the fine grid of points that cover the targets of interest. Let $\Delta$ denote the beam-width, then the desired beampattern at azimuth $\theta^\star$ is given by 	
\begin{equation}\label{beamwidth}
	\Phi(\theta)=\left\{ 
	\begin{aligned}
		1, & \kern 5pt \theta^\star-\frac{\Delta}{2} \le \theta \le \theta^\star + \frac{\Delta}{2} \\
		0, & \kern 5pt \text{otherwise}.
	\end{aligned}\right.
\end{equation}
Moreover, the second term is the mean-squared cross-correlation pattern, given by
\begin{equation} \label{CCC}
L_c(\textbf{R})=\frac{2}{P^2-P}\sum \limits_{p=1}^{P-1} \sum \limits_{q=p+1}^P |P_c(\bar{\theta}_p,\bar{\theta}_q;\textbf{R})|^2,
\end{equation} 
where $\{\theta_p\}_{p=1}^P$ are the given directions of the targets. We refer the reader to \cite{PS2007,LiuXiang2020} for more details.  
 
(2) \textit{\textbf{Performance metric for multi-user communication:}} The achievable transmission rate related to the SINR of the signal received by the downlink users is a standard performance measure in multiuser communication systems. For notation convenience, we introduce $\textbf{W}=[\textbf{W}_c, \textbf{W}_r]$, where $\textbf{w}_i$ is the $i$th column of $\textbf{W}$ for $i=1,\cdots,K+M$. Then, the signal covariance matrix can be rewritten as
\begin{equation} \label{rew}
\textbf{R}=\textbf{W}\textbf{W}^H=\sum\limits_{i=1}^{K+M}\textbf{w}_i\textbf{w}_i^H=\sum\limits_{i=1}^{K+M}\textbf{R}_i,
\end{equation}
where $\textbf{R}_i=\textbf{w}_i\textbf{w}_i^H$ is the rank 1 covariance matrix. Specifically, $\textbf{R}_1,\cdots,\textbf{R}_K$ are the covariance matrices of communication symbols, where the last $M$ ones are those of the radar waveforms. Thus, the SINR at the $k$th LU can be formulated as 
\begin{equation} \label{SINRbob}
\begin{aligned}
\gamma_k&=\frac{\mathbb{E}\{|\textbf{h}_k^H\textbf{w}_kc_k|^2 \}}{\sum\limits_{i=1, i\ne k}^{K}\mathbb{E}\{|\textbf{h}_k^H\textbf{w}_ic_i|^2 \}+\sum\limits_{j=1}^{M}\mathbb{E}\{|\textbf{h}_k^H\textbf{w}_{j+K}s_j|^2 \}+\sigma_c ^2} \\
&=\frac{\textbf{h}_k^H\textbf{R}_k\textbf{h}_k}{\sum\limits_{i=1, i\ne k}^{K+M}\textbf{h}_k^H\textbf{R}_i\textbf{h}_k+\sigma_c ^2}.
\end{aligned}
\end{equation}
There are two popular design criteria for multiuser beamforming \cite{Luo2011}. One of them is the throughput criterion to maximize the system's sum-rate. The other is the fairness criterion used for maximizing the minimal SINR at each user, which can be expressed as
\begin{equation}
\text{max} \kern 2pt \text{min}\{\gamma_1,\cdots,\gamma_K\}.
\end{equation} 
In this work, the SINR-fairness is adopted as the performance metric for multiuser communication. On the one hand, the fairness metric guarantees that each LU can obtain satisfactory QoS. On the other hand, the fairness metric based optimization is more tractable than the NP-hard optimal throughput beamforming problem. Given a minimal level of communication QoS $\Gamma_c$, the SNR-fairness metric can be transformed to forcing the minimal SINR of the users to be higher than the target threshold, i.e., $\gamma_k \ge \Gamma_c, k=1,\cdots,K$.

(3) \textbf{\textit{Performance metric for PLS level:}}
When the targets become Eves, the achievable data rates at Eves are non-negligible. A straightforward method is to increase the proportion of interference signal power to the detriment of the useful signal. According to the previous analysis, the radar waveform conveying no desired information can be regarded as the interference contaminating the reception of Eves. Accordingly, by recalling the received signal model (\ref{RecieveEve}), the SINR for the $q$th Eve can be formulated as 
\begin{equation} \label{SINREve}
\tilde{\gamma}_q=\frac{|\beta_q|^2\textbf{a}(\theta_q)^H\sum_{k=1}^{K}\textbf{R}_k\textbf{a}(\theta_q)}{|\beta_q|^2\textbf{a}(\theta_q)^H\sum_{i=K+1}^{K+M}\textbf{R}_i\textbf{a}(\theta_q)+\sigma_e ^2}.
\end{equation} 

Following \cite{NanSu2021}, we consider the worst-case SINR in (\ref{SINREve}), assuming that all the information intended for the $K$ LUs is the desired signal for Eves. As stated in \cite{ImCSI2011}, there will exist modulation and coding schemes that allow the LUs rather than the Eves to reliably decode the transmit information, as long as $\gamma_k > \tilde{\gamma}_q$, for $\forall k, q$. Therefore, we restrict the maximal SINR at Eves to be less than a given threshold $\Gamma_e$, instead of optimizing the secrecy rate $[\log(1+\gamma_k)-\log(1+\tilde{\gamma}_q)]^+$ defined in \cite{JC2018}, to achieve a satisfactory PLS level. On the one hand, the system's secrecy rate is difficult to determine due to its non-convexity with respect to $\textbf{R}_i$. On the other hand, since SINR-fairness based schemes are still capable of maintaining a minimal communication rate due to the monotonicity of the log function, we can equivalently achieve a desired secrecy rate $[\log(1+\Gamma_c)-\log(1+\Gamma_e)]^+$ by appropriately choosing the thresholds $\Gamma_c$ and $\Gamma_e$.

\section{The Beamforming Design for ideal scenarios}\label{JBCR}

In this section, we aim for designing the transmit beamforming matrices $\textbf{W}_r$ and $\textbf{W}_c$ under the consideration of the performance metrics for the radar beampattern, the communication QoS and the PLS levels given in the previous section. We first consider the ideal conditions, where the BS perfectly knows the CSI both for the LUs and Eves, and leave the beamformer design under the more practical imperfect CSI scenario for the next section.
  
\subsection{The proposed SDR-based beamforming algorithm}
Our beamforming design objective is to minimize the difference between the desired transmit beampattern and that generated by the BS to achieve good target detection and tracking performance. Meanwhile, the beamforming design also guarantees that the downlink SINR at the LUs remains higher than the given threshold, while that of the Eve is lower. Recalling the definition (\ref{rew}), instead of directly optimizing the precoding matrix $\textbf{W}$, the SDR based optimization problem with respect to the variables $\textbf{R}_i$ can be formulated as 
\begin{subequations}\label{Optimization0}
	\begin{align}
	\mathop { \text{minimize} }\limits_{\textbf{R},\{\textbf{R}_i\},\alpha} \kern 10pt & L_r(\textbf{R},\alpha) \kern 140pt (\mathcal{P}_0) \nonumber   \\
	\text{subject to} \kern 10pt &\textbf{R}=\sum\limits_{i=1}^{K+M} \textbf{R}_i \in \mathcal{S}_M^+, \kern 5pt \alpha>0, \kern 40pt  \label{a11}\\
	&\textbf{R}_i \in \mathcal{S}_M^+,\kern 2pt i=1,\cdots,K+M,  \label{b1}\\
    &\text{rank}(\textbf{R}_i)=1,\kern 2pt i=1,\cdots,K+M,  \label{b12}\\
	&[\textbf{R}]_{[m,m]}=P_t/M, \kern 2pt m=1,\cdots,M,  \label{c1}\\
	&\gamma_k \ge \Gamma_c, \kern 2pt k=1,\cdots,K,       \label{d1}      \\   
	&\tilde{\gamma}_q \le \Gamma_e, \kern 2pt q=1,\cdots,Q,  \label{f1}
	\end{align}
\end{subequations} 
where $\mathcal{S}_M^+$ represents the set consisting of all $M$-dimensional complex positive semidefinite matrices, i.e., $\mathcal{S}_M^+=\{\textbf{A}|\textbf{A} \in \mathbb{C}^{M \times M},\textbf{A}=\textbf{A}^H,\textbf{A} \succeq 0\}$. The rank-1 constraint in (\ref{b12}) is equivalent to $\textbf{R}_i=\textbf{w}_i\textbf{w}_i^H$. (\ref{c1}) represents the per-antenna power constraints, and $P_t$ is the total transmit power of the BS. Furthermore, the objective function and the constraints (\ref{d1}), (\ref{f1}) are the performance metrics introduced in Section \ref{PM}, where $\Gamma_c$ and $\Gamma_e$ are the predefined SINR thresholds at the LUs and Eve, respectively.

Upon substituting the SINR expressions (\ref{SINRbob}) as well as (\ref{SINREve}) into the constraints and applying some simple mathematical manipulations, (\ref{d1}) and (\ref{f1}) can be recast as
\begin{subequations} 
\begin{align}
(1+\Gamma_c^{-1})\textbf{h}_k^H\textbf{R}_k\textbf{h}_k \ge \textbf{h}_k^H\textbf{R}\textbf{h}_k+\sigma_c^2, \kern 2pt \forall k  \label{cCom}\\
(1+\Gamma_e^{-1})\textbf{a}_q^H\sum\limits_{k=1}^K\textbf{R}_k\textbf{a}_q \le \textbf{a}_q^H\textbf{R}\textbf{a}_q+\frac{\sigma_e^2}{|\beta_q|^2}, \forall q \label{cEve}
\end{align}
\end{subequations} 
where $\textbf{a}_q$ is the abbreviated form of $\textbf{a}(\theta_q)$. It can be observed that the individual matrices $\{\textbf{R}_i\}_{i\ge K+1}$ have no effect on the SINR constraints, which motivates us to remove these matrix variables from the original problem $\mathcal{P}_0$ of (\ref{Optimization0}). As a result, the number of matrix variables is reduced from $K+M+1$ to $K+1$, leading to much reduced memory requirements. By reformulating the constraint (\ref{a11}), problem $\mathcal{P}_0$ can be transformed to    
\begin{subequations}\label{Optimization1}
\begin{align}
\mathop { \text{minimize} }\limits_{\textbf{R},\textbf{R}_1,\cdots,\textbf{R}_K,\alpha} \kern10pt & L_r(\textbf{R},\alpha) \kern 135pt (\mathcal{P}_1) \nonumber  \\
\text{subject to} \kern 10pt &\textbf{R} \in \mathcal{S}_M^+, \kern 5pt \textbf{R}-\sum_{k=1}^K \textbf{R}_k \in \mathcal{S}_M^+,   \\
&\alpha>0,\kern 2pt \textbf{R}_k \in \mathcal{S}_M^+,\kern 2pt k=1,\cdots,K, \\
&\text{rank}(\textbf{R}_k)=1, \kern 2pt k=1,\cdots,K, \label{rank1p1}  \\
&[\textbf{R}]_{[m,m]}=P_t/M, \kern 2pt m=1,\cdots,M,  \\
&(\ref{cCom}), \kern 2pt (\ref{cEve}).   \nonumber
\end{align}
\end{subequations} 
However, problem $\mathcal{P}_1$ is non-convex due to the rank-1 constraints. Thus, the SDR relaxation based version of problem $\mathcal{P}_1$ can be obtained by omitting the rank-1 constraints (\ref{rank1p1}), which is denoted by problem $\mathcal{P}_2$. Thus, the problem $\mathcal{P}_2$ has become a standard quadratic semidefinite program (QSDP), since the objective function is a positive-semidefinite quadratic form and all the constraints are either linear or semidefinite. Hence, the global optimum can be obtained in polynomial time with the aid of standard convex optimization toolboxes \cite{CVX1,sdp3}. Note that the optimal solutions of the relaxed problem $\mathcal{P}_2$ are not necessarily rank-1 matrices, hence either the classic eigenvalue decomposition or Gaussian randomization methods \cite{Luo2010} can be leveraged to obtain the solutions of the original problem $\mathcal{P}_1$. Unfortunately, these kinds of approximate algorithms usually only provide suboptimal solutions of the original problem, hence resulting in a loss of performance.
    
To circumvent this deficiency, we set out to find a global optimum for problem $\mathcal{P}_1$, which means that the SDR relaxation is tight. Inspired by the result in \cite{LiuXiang2020}, we propose the following proposition. 

\textit{Proposition 1}: Let $\hat{\textbf{R}},\hat{\textbf{R}}_1,\cdots,\hat{\textbf{R}}_K$ be the optimal solution of the QSDP problem $\mathcal{P}_2$. There also exists a global optimum $\tilde{\textbf{R}},\tilde{\textbf{R}}_1,\cdots,\tilde{\textbf{R}}_K$ for problem $\mathcal{P}_1$, where we have   
\begin{equation} \label{optimal}
\tilde{\textbf{R}}=\hat{\textbf{R}},\kern 2pt \tilde{\textbf{w}}_k=(\textbf{h}_k^H\hat{\textbf{R}}_k\textbf{h}_k)^{-1/2}\hat{\textbf{R}}_k\textbf{h}_k, \kern 2pt \tilde{\textbf{R}}_k=\tilde{\textbf{w}}_k\tilde{\textbf{w}}_k^H,
\end{equation}
for $k=1,\cdots,K$.

\textit{Proof}: The proof is relegated to Appendix \ref{A}. $\hfill\blacksquare$

According to Proposition 1, we can get the global rank-1 optimal solution for problem $\mathcal{P}_1$ from its QSDP relaxation based version $\mathcal{P}_2$, where the relaxation is tight. The remaining step is to find the optimal solution for the original problem $\mathcal{P}_0$, i.e. obtaining the precoding matrix $\textbf{W}_r$ for the radar waveforms. To meet the constrains of (\ref{a11}) and (\ref{b1}), the $M$ precoding vectors $\{\textbf{w}_i\}_{i \ge K+1}$ can be obtained by the following decomposition
\begin{equation}\label{decomposition} 
\textbf{W}_r\textbf{W}_r^H=\textbf{R}_\text{rad}=\tilde{\textbf{R}}-\sum\limits_{k=1}^K\tilde{\textbf{R}}_k,
\end{equation}
where $\textbf{W}_r=[\textbf{w}_{K+1},\cdots,\textbf{w}_{K+M}]$. Actually, since the associated waveform level design is not considered in this work, the decomposition (\ref{decomposition}) is not unique, but it is trivial thanks to the positive semi-definite nature of the radar signal's covariance matrix. Several decomposition methods such as the square root matrix ($\textbf{W}_r=\textbf{R}_\text{rad}^{\frac{1}{2}}\textbf{U}$, $\textbf{U}$ is an arbitrary unitary matrix) based one \cite{PSLJ2008} and the Cholesky decomposition based one may be applied \cite{ZXM}. 
 
%

\subsection{The ZF-based low complexity algorithm}\label{ZFlowComplexity}

The main computational complexity burden in the proposed SDR-based algorithm is imposed by that of solving the QSDP problem $\mathcal{P}_2$, which motivates us to seek a low-complexity solution. Inspired by the zero forcing (ZF) based method of \cite{LiuXiang2020}, we develop a reduced-complexity sub-optimal algorithm by incorporating ZF constraints into problem $\mathcal{P}_2$. The ZF method is widely used in low-complexity linear precoders, because its performance tends to that of the optimal non-linear precoder, especially for a large number of antennas \cite{ALS2004,AW2008}. Its main appeal is that of eliminating the inter-user and radar interferences, hence achieving a high SINR at each user. Mathematically, the ZF constraints can be expressed as
\begin{equation}\label{ZF} 
	\textbf{H}\textbf{W}_c=\text{diag}(\sqrt{\rho_1},\cdots,\sqrt{\rho_K}), \kern 2pt \textbf{H}\textbf{W}_r=\textbf{0}_{K \times M},
\end{equation}     
where $\rho_k$ represents the signal power at the $k$th user, for $1 \le k \le K$. Upon recalling the definition $\textbf{W}=[\textbf{W}_c,\textbf{W}_r]$ and $\textbf{R}=\textbf{W}\textbf{W}^H$, (\ref{ZF}) can be equivalently transformed to the following form (Theorem 2,\cite{LiuXiang2020})
\begin{equation}\label{ZFF} 
	\textbf{H}\textbf{R}\textbf{H}^H=\text{diag}(\bm{\rho}),
\end{equation}
where $\bm{\rho}=[\rho_1,\cdots,\rho_K]$. Moreover, substituting (\ref{ZF}) or (\ref{ZFF}) into the SINR expression (\ref{SINRbob}), the associated SINR constraints (\ref{cCom}) can be simplified by
\begin{equation}\label{ZFCom}
\rho_k \ge \Gamma_c \sigma_c^2, \kern 2pt \forall k.
\end{equation}
It can be observed that the individual matrix variable $\textbf{R}_k$ has been removed from the SINR constraints for the LUs by imposing the ZF constraints. Following the same methodology for further reducing the number of matrix variables, and by introducing the auxiliary matrix variable $\textbf{R}_\text{com}=\sum_{k=1}^K\textbf{R}_k$, the PLS constraint (\ref{cEve}) can be rewritten as follows  
\begin{equation}\label{ZFEve} 
(1+\Gamma_e^{-1})\textbf{a}_q^H\textbf{R}_\text{com}\textbf{a}_q \le \textbf{a}_q^H\textbf{R}\textbf{a}_q+\frac{\sigma_e^2}{|\beta_q|^2}, \kern 2pt \forall q
\end{equation}  
Furthermore, we can immediately infer the ZF constraint for $\textbf{R}_\text{com}$ as  
\begin{equation}\label{ZFFC} 
	\textbf{H}\textbf{R}_\text{com}\textbf{H}^H=\textbf{H}\textbf{W}_c\textbf{W}_c^H\textbf{H}^H=\text{diag}(\bm{\rho}).
\end{equation}
As a consequence, either the communication SINR constraint or the PLS constraint no longer contains the individual matrix variable $\textbf{R}_k$. Accordingly, problem $\mathcal{P}_2$ can be converted to
\begin{subequations}\label{OptimizationZF}
	\begin{align}
		\mathop { \text{minimize} }  \limits_{\textbf{R},\textbf{R}_\text{com},\bm{\rho},\alpha} \kern 10pt & L_r(\textbf{R},\alpha) \kern 135pt (\mathcal{P}_3) \nonumber   \\
		\text{subject to} \kern 10pt & \textbf{R} \in \mathcal{S}_M^+,{\kern 5pt} \textbf{R}-\textbf{R}_\text{com} \in \mathcal{S}_M^+, \textbf{R}_\text{com} \in \mathcal{S}_M^+, \label{ZFs}   \\
		&[\textbf{R}]_{[m,m]}=P_t/M, \kern 2pt m=1,2,\cdots,M,  \\
		&\textbf{H}\textbf{R}\textbf{H}^H=\text{diag}(\bm{\rho}), \\
		&\textbf{H}\textbf{R}_\text{com}\textbf{H}^H=\text{diag}(\bm{\rho}),\\
		&\alpha>0,\kern 2pt (\ref{ZFCom}), \kern 2pt (\ref{ZFEve}). \nonumber
	\end{align}
\end{subequations} 
Problem $\mathcal{P}_3$ is also a standard QSDP problem, because the objective function has a positive-semidefinite quadratic form and all the constraints are either linear or semidefinite. Similarly, the optimal solutions $\hat{\textbf{R}}$ and $\hat{\textbf{R}}_\text{com}$ can be obtained by a standard convex optimization toolbox in polynomial time. 

The next step is to recover the precoding matrix $\textbf{W}$ from the optimal solutions $\hat{\textbf{R}}$ and $\hat{\textbf{R}}_\text{com}$. Inspired by Theorem 2 of \cite{LiuXiang2020}, we conceive the following procedure of constructing the radar and communication precoding matrices, respectively. First, either the classic Cholesky decomposition or square root method is used by exploiting the positive-semidefinite property for $\hat{\textbf{R}}_\text{com}=\textbf{L}_c\textbf{L}_c^H$. Then, we employ the row QR decomposition of $\textbf{H}\textbf{L}_c$, yielding
\begin{equation}\label{QR} 
\textbf{H}\textbf{L}_c=[\textbf{L}_h,\textbf{0}_{K \times (M-K)}]\textbf{Q},
\end{equation}
where $\textbf{L}_h$ is a $K \times K$ lower triangular matrix and $\textbf{Q}$ is a $M \times M$ unitary matrix. Thus, the communication precoder can be formulated as   
\begin{equation}\label{Wc} 
\textbf{W}_c=\textbf{L}_c[\textbf{Q}^H]_{[:,1:K]},
\end{equation}
while the radar precoding matrix $\textbf{W}_r$ can be expressed as
\begin{equation}\label{WrO} 
\textbf{W}_r\textbf{W}_r^H=\hat{\textbf{R}}_\text{rad}=\hat{\textbf{R}}-\textbf{W}_c\textbf{W}_c^H.
\end{equation} 
Subsequently, we analyze the feasibility of the proposed precoder design method by introducing the following proposition. 

\textit{Proposition 2}: Given the optimal solution $\hat{\textbf{R}}$ and $\hat{\textbf{R}}_\text{com}$ of problem $\mathcal{P}_3$, the matrices $\textbf{W}_c$ in (\ref{Wc}) and $\textbf{W}_r$ in (\ref{WrO}) are also the optimal precoders of problem $\mathcal{P}_3$ and satisfy the ZF constraint (\ref{ZF}) at the same time. 

\textit{Proof}: The proof is divided into three parts, and it is relegated to Appendix \ref{B}.$\hfill\blacksquare$

Proposition 2 illustrates the feasibility and efficiency of the proposed precoding matrices recovered from the optimal solution of problem $\mathcal{P}_3$. In summary, we can obtain the optimal beamforming for DFRC secure transmission with the perfectly known CSI by the proposed SDR-based and the low complexity ZF-based algorithms. The detailed procedure of the proposed algorithms are summarized in Algorithm \ref{alg}.
\begin{algorithm}[!t]
	\caption{ The proposed SDR(ZF)-based beamforming algorithm designed for secure DFRC.}   
	\label{alg}
	\begin{algorithmic}		
		\STATE \textbf{Input:} \\
		Total transmit power of base station $P_t$;\\
		Radar desired beampattern $\Phi(\theta)$; \\
		Instantaneous downlink channel $\textbf{H}$; \\
		SINR threshold at LUs $\Gamma_c$ and at Eves $\Gamma_e$; \\
		The directions of Eves $\theta_q, q=1,\cdots,Q$;
		\STATE \textbf{Output} \\ 
		The overall precoding matrix $\textbf{W}=[\textbf{w}_1,\cdots,\textbf{w}_{K+M}]$.
		\STATE \textbf{Steps} \\
		1. Compute the optimal solution of $\mathcal{P}_2$ (or $\mathcal{P}_3$) via convex optimization solver;  \\
		2. Compute $\textbf{w}_1,\cdots,\textbf{w}_K$ by (\ref{optimal}) (or by (\ref{Wc}));  \\
		3. Compute $\textbf{w}_{K+1},\cdots,\textbf{w}_{K+M}$ by (\ref{decomposition}) (or by (\ref{WrO})); \\ 		
	\end{algorithmic}
\end{algorithm}

\section{Robust Beamforming design with imperfect CSI knowledge} \label{robust}
In practice, it is challenging to obtain the exact CSI due to the estimation errors, feedback quantization, hardware deficiencies, etc., resulting in imperfect CSI knowledge at the BS. Specifically, for the radar targets, we assume that the direction of the $q$-th target is roughly known by the BS within an angular interval of $[\theta_q-\Delta \theta_q,\theta_q+\Delta \theta_q]$, where $\Delta \theta_q$ represents the associated angle uncertainty. Moreover, for the communication LUs, the additive error model of the CSI matrix for the $k$-th LU is considered as $\textbf{h}_k = \hat{\textbf{h}}_k+\bm{\epsilon}_k$, where $\hat{\textbf{h}}_k$ is the estimated CSI matrix and $\bm{\epsilon}_k$ denotes the channel uncertainty. To this end, we aim for designing the robust beamforming scheme for secure transmission in this section. 

\subsection{Wide main-lobe beampattern design}
The uncertainties of the target directions have an impact on both the objective function and the PLS constraints in problem $\mathcal{P}_0$. On one hand, the BS should form a wide main-lobe to avoid missing the target. Thus, the beam-width $\Delta$ in (\ref{beamwidth}) should be appropriately chosen according to the angular uncertainty $\Delta \theta_q$, in order to cover all the possible locations of the target.

On the other hand, since Eve may be located in an arbitrary direction within the angular interval, we should guarantee a satisfactory secrecy rate for every possible direction. Consequently, the SINR constraints (\ref{cEve}) should be modified according to   
\begin{equation}\label{Uncertain} 
(1+\Gamma_e^{-1})\textbf{a}^H_{q_i}\sum\limits_{k=1}^K\textbf{R}_k\textbf{a}_{q_i} \le \textbf{a}^H_{q_i}\textbf{R}\textbf{a}_{q_i}+\frac{\sigma_e^2}{|\beta_q|^2}, \kern 2pt \forall \theta_{q_i} \in \bar{\Omega}_q,
\end{equation}
where $\bar{\Omega}_q$ is a discrete set that covers the potential directions of the $q$-th Eve, and $\textbf{a}_{q_i}$ represents the compact form of $\textbf{a}(\theta_{q_i})$. It can be observed that the angular uncertainty introduces more constraints similar to (\ref{cEve}) over the associated angular interval. Evidently, the proposed Algorithm \ref{alg} is also capable of handling the modified constraints (\ref{Uncertain}). In other words, the number of targets and the uncertainty of target directions determine the number of PLS constraints. Naturally, imposing a large number of constraints for securing certain PLS levels results in degraded radar beampattern and communication QoS. We will illustrate this phenomenon in Section \ref{Simulation}.    

\subsection{Robust beamforming for mitigating CSI error of LUs }

Similar to \cite{NanSu2021,Imperfect2014}, we assume that the CSI uncertainty is bounded by a spherical region as
\begin{equation} \label{e1}
	\mathcal{S}_k := \{ \hat{\textbf{h}}_k+\bm{\epsilon}_k \kern 2pt | \kern 2pt  || \bm{\epsilon}_k|| \le u_k \}, \kern 2pt \forall k.
\end{equation}
In this case, the SINR expression for the $k$-th LU in (\ref{SINRbob}) should be replaced by the worst-case SINR over the set $\mathcal{S}_k$, namely
\begin{equation} \label{e2}
	\bar{\gamma}_k = \mathop { \text{min} }  \limits_{ \textbf{h}_k \in \mathcal{S}_k } \gamma_k, \kern 2pt \forall k. 
\end{equation}
Thus, based on the definitions (\ref{e1}) and (\ref{e2}), the SINR constraint in (\ref{cCom}) can be reformulated as
\begin{equation} \label{robustSINR}
(\hat{\textbf{h}}_k+\bm{\epsilon}_k)^H \left[ (1+\Gamma_c^{-1})\textbf{R}_k-\textbf{R} \right] (\hat{\textbf{h}}_k+\bm{\epsilon}_k)-\sigma_c^2 \ge 0, \kern 2pt \forall k.
\end{equation}
Then, we adopt the popular S-procedure of robust optimization to tackle the SINR constraints mentioned above. By introducing an auxiliary vector $\textbf{t}=[t_1,\cdots, t_K]$, the original problem $\mathcal{P}_1$ can be reformulated as the following robust beamforming version  \cite{NanSu2021,Imperfect2014} 
\begin{equation} 
	\begin{aligned}
		\mathop { \text{minimize} }\limits_{\textbf{R},\textbf{R}_1,\cdots,\textbf{R}_K, \textbf{t}, \alpha} \kern10pt  L_r(\textbf{R},\alpha) \kern 130pt (\mathcal{P}_4)   \\
		\text{subject to} \kern 10pt  (\ref{Optimization1}a) - (\ref{Optimization1}d), \kern 2pt (\ref{cEve}) \kern 2pt \text{or} \kern 2pt (\ref{Uncertain}), \kern 75pt   \\
		\left( {\begin{array}{*{20}{c}}
					{\textbf{S}_k + t_k \textbf{I}_M}&{ \textbf{S}_k \hat{\textbf{h}}_k }\\
					{\hat{\textbf{h}}_k^H \textbf{S}_k}&{ \textbf{h}_k^H \textbf{S}_k \textbf{h}_k - \sigma_c^2 - t_k u_k^2   }
			\end{array}} \right) \succeq 0, \forall k \\
		\textbf{S}_k : = (1+\Gamma_c^{-1})\textbf{R}_k-\textbf{R}, \kern 2pt t_k \ge 0. \kern 60pt
	\end{aligned}
\end{equation} 
Again, by dropping the rank-1 constraints (\ref{rank1p1}), problem $\mathcal{P}_4$ becomes a QSDP, which can be efficiently solved in polynomial time. Then, we will show that the optimal solution of the QSDP reconstruction method in (\ref{optimal}) also holds for the proposed robust beamforming.      

\textit{Proposition 3}:  Let $\hat{\textbf{R}},\hat{\textbf{R}}_1,\cdots,\hat{\textbf{R}}_K$ be the optimal solution of the relaxed version of problem $\mathcal{P}_4$. Then the $\tilde{\textbf{R}},\tilde{\textbf{R}}_1,\cdots,\tilde{\textbf{R}}_K$ associated with the expression of (\ref{optimal}) is also the optimal solution of the original problem $\mathcal{P}_4$.

\textit{Proof}: By employing the result in Proposition 1, the proof becomes straightforward upon substituting (\ref{optimal}) into the constraints (\ref{robustSINR}). $\hfill\blacksquare$

\section{Performance And Complexity Analysis}\label{USS}
\subsection{Complexity Analysis}\label{CA}
The complexity of the proposed algorithms is dominated by the QSDP problem. For a given solution accuracy $\epsilon$, the worst-case complexity order of solving problem $\mathcal{P}_2$ using the primal-dual interior-point algorithm is $\mathcal{O}[(K+Q)^{6.5}M^{6.5}\text{log}(1/\epsilon)]$ \cite{ComQSDP,LiuXiang2020}, where $K+Q$ and $M$ refer to the number of semidefinite constraints and the dimension of matrix variables, respectively. Compared to the SDR algorithm, the low complexity ZF beamforming problem $\mathcal{P}_3$ includes $5=\mathcal{O}(1)$ such constraints, hence the worst-case complexity order becomes $\mathcal{O}[Q^{6.5}M^{6.5}\text{log}(1/\epsilon)]$. Furthermore, for the robust beamforming algorithm with imperfect CSI knowledge, the complexity also depends on the number of elements in the set $\bar{\Omega}_q$ of (\ref{Uncertain}). Specifically, upon denoting the cardinality of the set $\bar{\Omega}_q$ as $P$, the worst-case complexity is on the order of $\mathcal{O}[K^{6.5}P^{6.5}M^{6.5}\text{log}(1/\epsilon)]$.          

\subsection{Performance Analysis} \label{PAnalysis}
In this subsection, we provide the performance analysis of the proposed algorithms.

(I) We can immediately spot the performance trade-off among the radar beampattern, the communication QoS, and the PLS level in problem $\mathcal{P}_1$. The constraints (\ref{cCom}) and (\ref{cEve}) always hold, when we have $\Gamma_c=0$ and $\Gamma_e \to \infty $. In this case, problem $\mathcal{P}_1$ is reduced to the conventional radar-only beamforming design, determining the optimal beampattern for radar detection. Explicitly, any improvements of the communication QoS and PLS level are attained at the cost of sacrificing the radar performance, since the radar loss function will increase upon increasing $\Gamma_c$ or decreasing $\Gamma_e$.

(II) Compared to the SDR-based algorithm, the low complexity ZF-based algorithm forces the radar and inter-user interference to zero, potentially raising the SINR at the LUs to a certain threshold (denoted by $\hat{\Gamma}$). Thus, for the communication constraints, we have
	\begin{equation} \label{SNRr}
		\left\{
		\begin{aligned}
			\gamma_k^\text{ZF}  = \hat{\Gamma} > \gamma_k^\text{SDR} \ge \Gamma_c, \kern 5pt \text{when} \kern 2pt \Gamma_c < \hat{\Gamma}, \\
			\gamma_k^\text{ZF}  = \gamma_k^\text{SDR} \ge \Gamma_c \ge \hat{\Gamma} , \kern 5pt \text{when} \kern 2pt \Gamma_c \ge \hat{\Gamma}.
		\end{aligned}
		\right. \tag{35}
	\end{equation}
For a relatively low threshold $\Gamma_c$, the interference encountered by the users do not have to be as low as zero to satisfy the SINR constraint, resulting in $\gamma_k^\text{ZF} > \gamma_k^\text{SDR}$. By contrast, the interference in $\gamma_k^\text{SDR}$ has to be eliminated to meet the high SINR requirements, resulting in $\gamma_k^\text{ZF} = \gamma_k^\text{SDR}$. According to (35), we can immediately conclude the following properties of the ZF-based algorithm. (1) It results in worse radar beampattern than the SDR-based algorithm because more severe restrictions are imposed by the ZF constraint when $\Gamma_c < \hat{\Gamma}$. (2) The radar loss function and the users' SINR remains constant, as long as the threshold $\Gamma_c$ is lower than a positive value $\hat{\Gamma}$. (3) The performance of ZF-based beamforming tends to be similar to that of SDR-based beamforming at high SINRs, i.e., $\Gamma_c \ge \hat{\Gamma}$.  

(III) For the SDR-based algorithm, the system's secrecy rate is always approximated by $\text{log}_2(1+\Gamma_c)-\text{log}_2(1+\Gamma_e)$ given the thresholds $\Gamma_c$ and $\Gamma_e$, because the optimal solution generally reaches the boundary of the feasible region. By contrast, the secrecy rate of the ZF-based algorithm may become higher than the above value for small $\Gamma_c$ values due to the potentially high SINR achieved under the ZF constraint. The proposed algorithms guarantee to have a secrecy rate above a certain lower bound. 

(IV) Upon considering the extreme case that the channels of the users and Eves have the same quality, i.e., $\beta_k\textbf{a}(\theta_k)=\textbf{h}_k$, the communication QoS constraint (\ref{cCom}) and the PLS level constraint (\ref{cEve}) are contradictory to each other, hence leading to the infeasibility of problem $\mathcal{P}_1$. This means that the feasibility probability of problem $\mathcal{P}_1$ significantly depends on the values of $\Gamma_c$ as well as $\Gamma_e$, and on the distances between the targets and Eves. The proposed joint beamforming design method will become invalid, when the Eves are at the same directions as the users. The symbol-level range sidelobe design \cite{Sidelobe2020} may be a promising remedy, which we will leave for future research.       

(V) It should be pointed out that the joint PLS beamforming design of \cite{NanSu2021} minimized the SINR at Eve, which is different from the proposed method optimizing the radar transmit beampattern. Even though it cannot be compared directly due to the different functional requirements, the proposed method has the following advantages over \cite{NanSu2021}. (1) The fractional programming approach is adopted in \cite{NanSu2021}, where a sequence of SDPs has to be solved by iteration, imposing a heavy computational burden. By contrast, the proposed methods only have to solve a SDP or QSDP problem with the same number of matrix variables. (2) The eigenvalue decomposition or Gaussian randomization techniques of \cite{NanSu2021} result in a sub-optimal solution, when the ranks of the optimal matrices obtained by the SDP solver are not equal to 1. By contrast, the proposed SDR relaxation is tight. (3) When using the SINR instead of the secrecy rate as the objective, the difference between the achievable rate of users and that of Eves may become negative, leading to a secrecy rate of $\text{SR}=0$. By contrast, the proposed algorithms can always guarantee a satisfactory secrecy rate.

\section{Simulation Results}\label{Simulation}

In this section, we evaluate the proposed joint PLS beamforming algorithm by numerical simulations. The system parameters are set as follows, unless specified otherwise. The BS is equipped with a ULA having half-wavelength spacing between adjacent antennas, i.e. $d/\lambda=1/2$. The number of antennas is set to $M=10$, and the total transmit power is normalized as $P_t=1$. The angular directions are obtained by uniform sampling with resolution of $0.1^\circ$, including $\{\theta_l\}_{l=1}^L$ in (\ref{DDD}) with the range of $[-90^\circ,90^\circ]$, and $\Omega_q$ in (\ref{Uncertain}). Without loss of generality, we adopt the Rayleigh fading model for the multi-user communication channel so that each entry of $\textbf{H}$ obeys the standard complex Gaussian distribution with $h_{i,j}\sim \mathcal{CN}(0,1)$. Additionally, we assume the noise levels at the Eves and LUs to be the same, i.e., $\sigma_c^2=\sigma_e^2=0.01$ for convenience. The individual radar waveforms and communication symbols are generated as random quadrature-phase-shift-keying (QPSK) modulated sequences, with the total number of symbols being $N=1024$.

For comparison, we choose the joint beamforming design method  and its low-complexity counterpart proposed in \cite{LiuXiang2020} termed as \textit{Benchmark 1} and \textit{Benchmark 2}, respectively. Compared to \cite{LiuFan2018}, where only the communication signal is exploited by the DFRC system, the superiority of the combined radar waveforms and communication signals in terms of increasing the DoFs has been shown in \cite{LiuXiang2020}. Therefore, we refer to \cite{LiuXiang2020} for circumventing repetition.
  
First, we numerically characterize the MIMO radar transmit beampattern, where the proposed SDR-based algorithm and its low-complexity version are referred to as SDR and ZF, respectively. We set the direction of a single target to $\theta_0=0^\circ$, the threshold for the LUs' SINRs to $\Gamma_c=10$dB, and the threshold for the Eve's SINR to $\Gamma_e=0$dB. Fig. \ref{Compare_Pattern} illustrates the trade-off among the radar beampattern, the communication QoS and the PLS level. Although the proposed algorithms impose a performance degradation on the transmit beampattern compared to their counterparts, the target secrecy rate (SR) can still be guaranteed. By contrast, \textit{Benchmark} 1 and 2 form better beampatterns, but their SR becomes zero. Then, we evaluate the system performance versus the predefined SINR thresholds $\Gamma_c$ and $\Gamma_e$, respectively.

\subsection{System Performance Evaluation vs. the Threshold $\Gamma_c$}\label{SingleT_Sim}

 \begin{figure}[!t]
	\centering
	\includegraphics[width=3in]{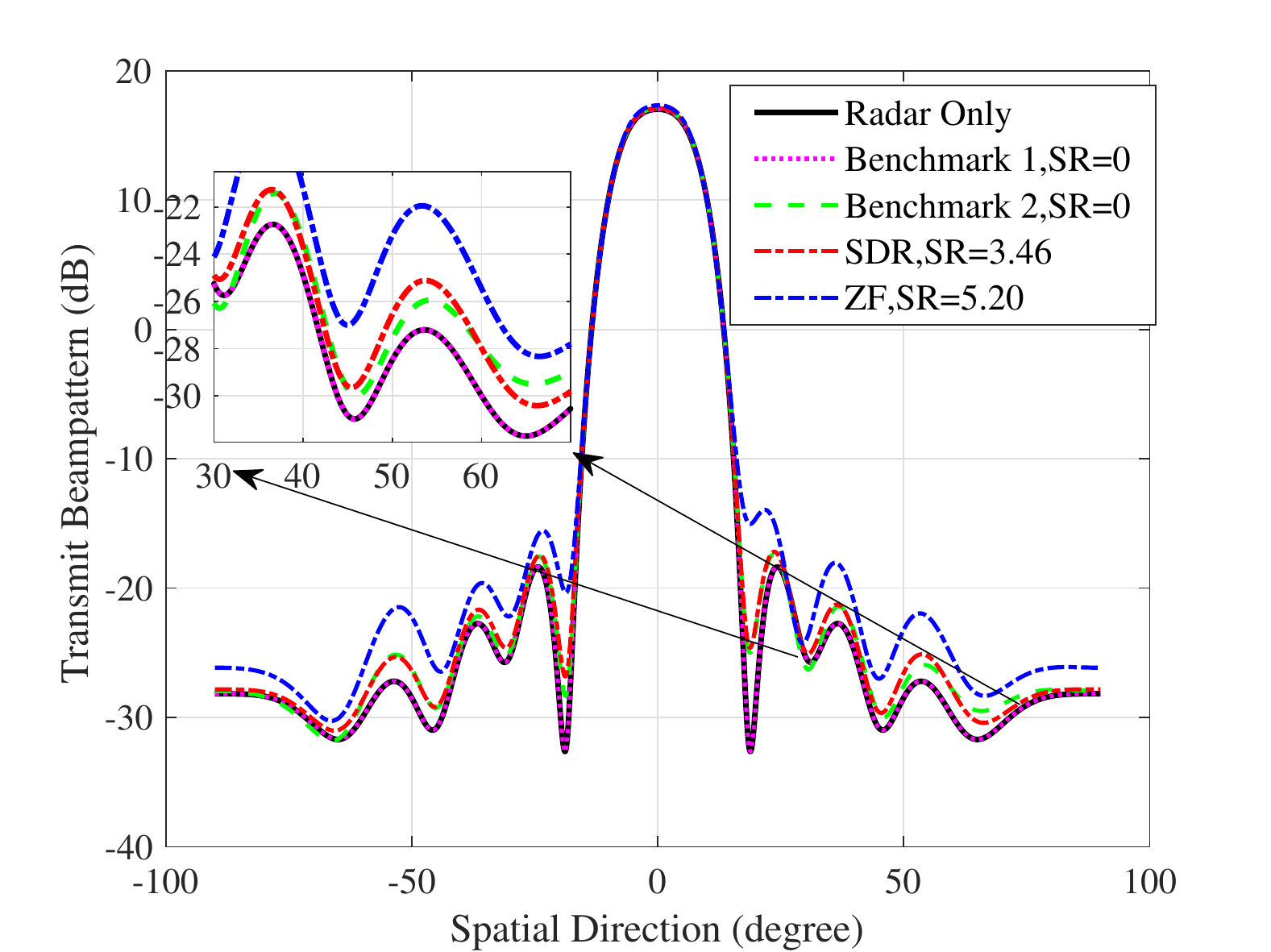}
	\caption{Radar transmit beampattern for the direction $\theta_0=0^\circ$, with $K=2$, $\Gamma_c=10$dB, and $\Gamma_e=0$dB.}
	\label{Compare_Pattern}
	\setcounter{figure}{2}
\end{figure}

In this subsection, we keep the SINR threshold of Eves $\Gamma_e=0$dB as a constant, and sweep $\Gamma_c$ of LUs from $10$dB to $18$dB to test its impact. All of the simulation results represent averaged values over 500 Monte Carlo trials. In each trial, the target direction $\theta_q$ is chosen randomly in the range of $[-60^\circ,60^\circ]$, and the CSI of the link spanning from the BS and the LUs obey the standard Complex Gaussian distribution. The radar performance is evaluated as the difference between the DFRC transmit beampattern and the optimal radar-only beampattern by defining the mean square error (MSE) metric as  
\begin{equation}\label{MSE} 
\text{MSE}=\frac{1}{L}\sum \limits_{l=1}^L |P(\theta_l;\hat{\textbf{R}})-P(\theta_l;\textbf{R}^\star)|^2,
\end{equation}  
where $\textbf{R}^\star$ is the optimal radar-only variance matrix by the 3dB low sidelobe beampattern design scheme of \cite{PS2007} .

\begin{figure*}[htbp]	
	\subfigure{
	\begin{minipage}[t]{0.3\linewidth}
			\centering
			\includegraphics[width=2.3in]{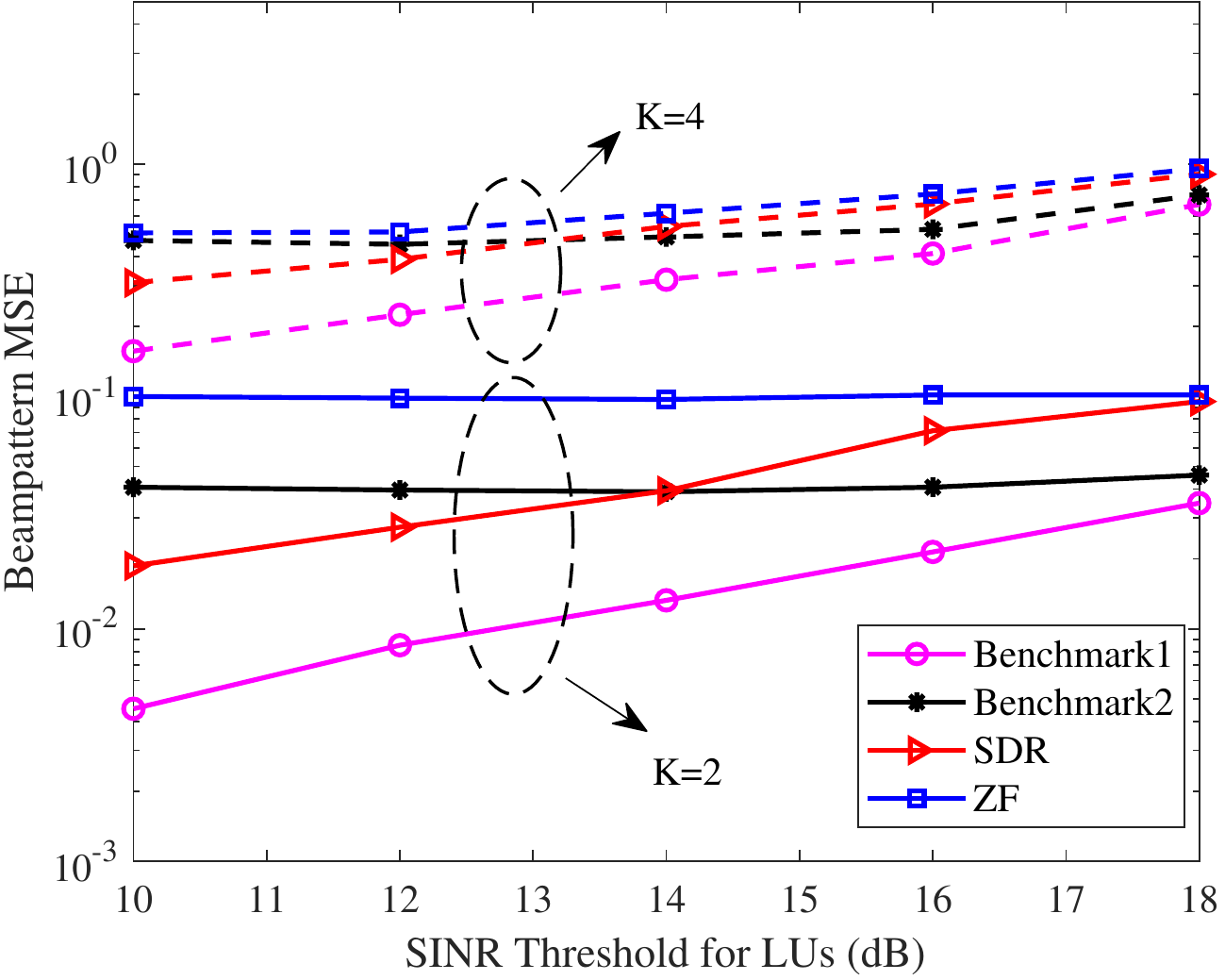}
			\caption{Beampattern MSE versus SINR threshold $\Gamma_c$ for LUs, $\Gamma_e=0$dB.}
			\label{Compare_BobTs_MSE}
	\end{minipage}}
	\hspace{0.01\linewidth}	
	\subfigure{
	\begin{minipage}[t]{0.3\linewidth}
			\centering
			\includegraphics[width=2.3in]{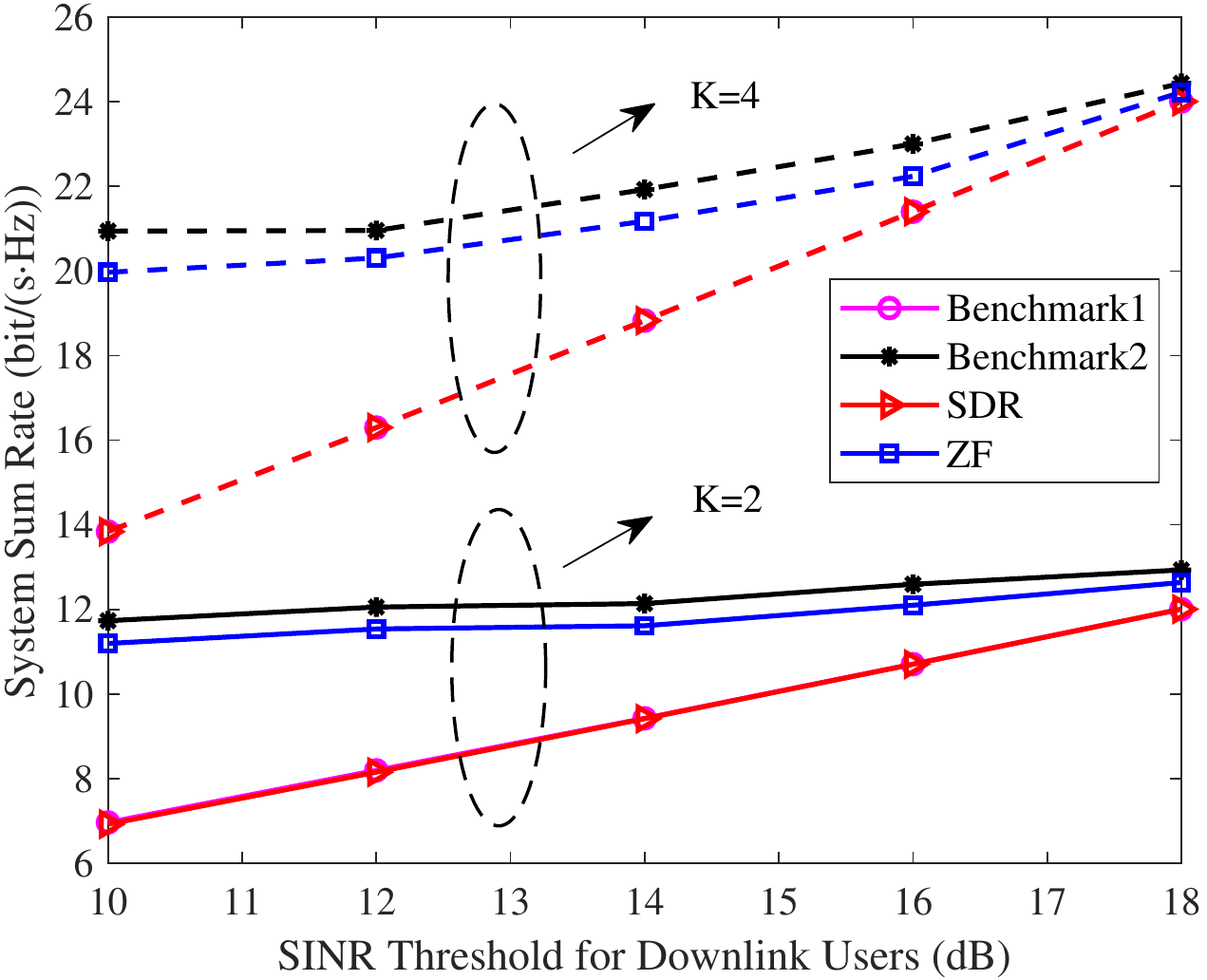}
			\caption{The achievable sum rate versus SINR threshold $\Gamma_c$ for LUs, $\Gamma_e=0$dB.}
			\label{Compare_BobTs_SumRate}
	\end{minipage}}
	\hspace{0.01\linewidth} 
	\subfigure{
	\begin{minipage}[t]{0.3\linewidth}
			\centering
			\includegraphics[width=2.3in]{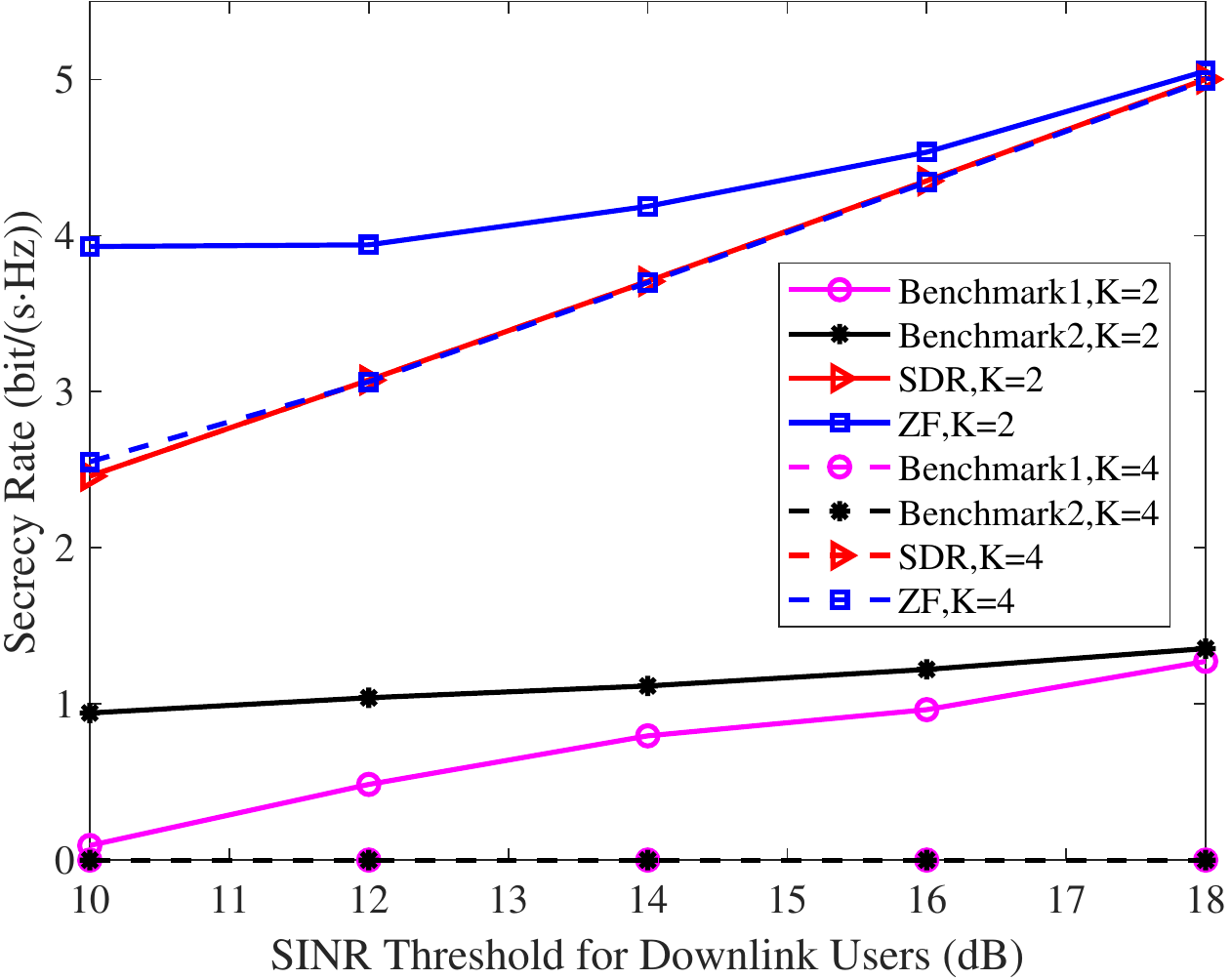}
			\caption{The secrecy rate versus SINR threshold $\Gamma_c$ for LUs, $\Gamma_e=0$dB.}
			\label{Compare_BobTs_PLS}
	\end{minipage}}
\end{figure*}

Fig. \ref{Compare_BobTs_MSE} shows the beampattern MSE versus the SINR threshold $\Gamma_c$ of the LUs. We can observe the following three phenomena from Fig. \ref{Compare_BobTs_MSE}. (1) The beampattern MSEs of all algorithms increase upon increasing $\Gamma_c$, which is consistent with the previous analysis. As expected, the MSE of the ZF-based algorithms remains constant in the scenarios of $K=2$ and for SINRs below 16dB at $K=4$. This is because the ZF-based methods force the interference to zero, leading to a potentially high SINR. Thus, the performance will remain constant until the SINR thresholds become higher than the potential SINR achieved by the ZF constraint. The performance gaps between the SDR-based and ZF-based methods become quite small for high enough values of $\Gamma_c$. (2) The benchmark algorithms formulate better beampattern, since the PLS aspects of confidential information protection is not taken into account in these methods. (3) The more users have to be supported, the higher the beampattern MSE becomes. Notably, the impact of the number of users $K$ on the beampattern MSE is more significant than that of the SINR threshold $\Gamma_c$, which implies that serving more downlink users is more restrictive than improving the SINR level of the users.


\begin{figure*}[htbp]	
	\subfigure{
		\begin{minipage}[t]{0.3\linewidth}
			\centering
			\includegraphics[width=2.3in]{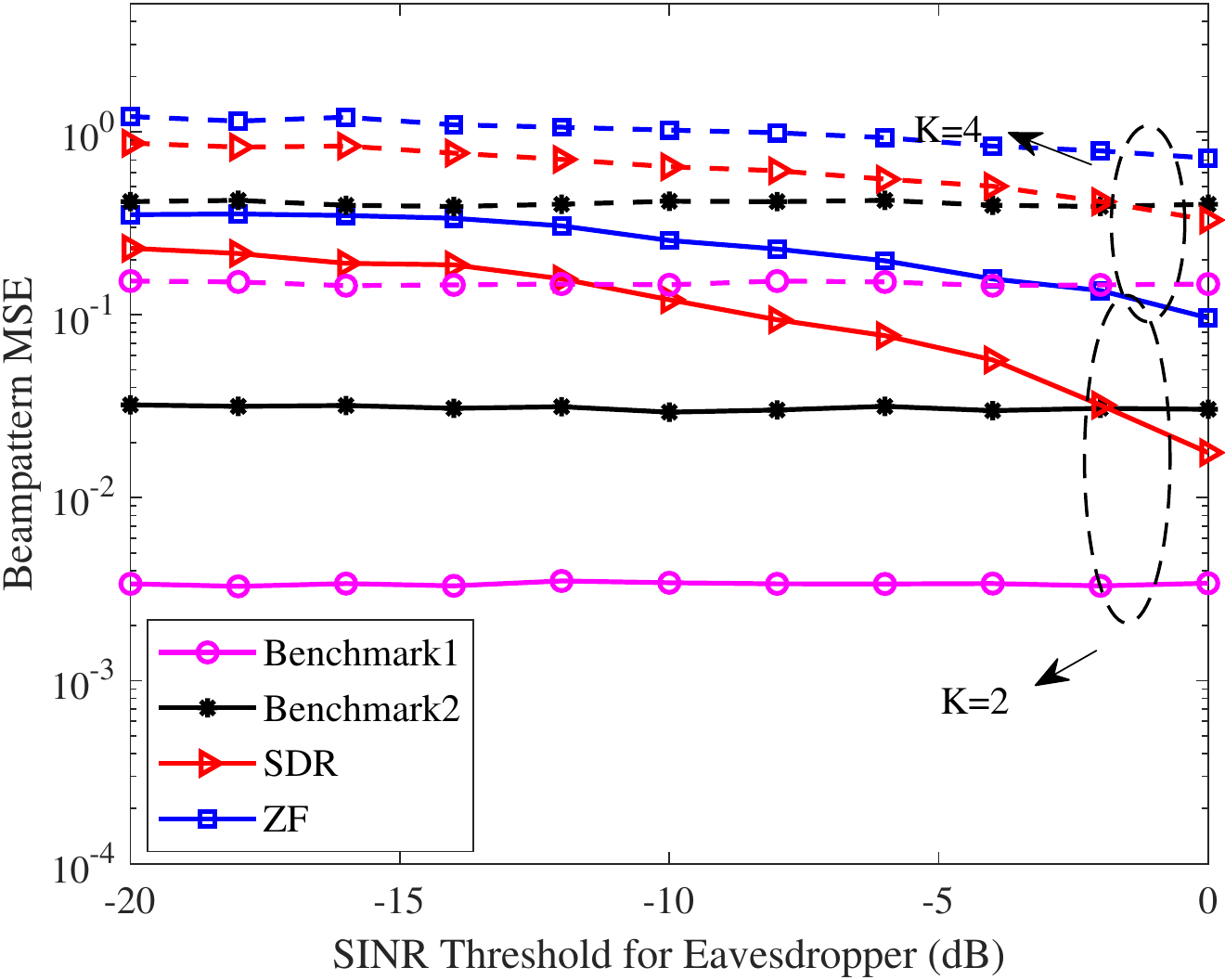}
			\caption{Beampattern MSE versus SINR threshold $\Gamma_e$ for Eves, $\Gamma_c=10$dB.}
			\label{Compare_Eve_MSE}
	\end{minipage}}
	\hspace{0.01\linewidth}	
	\subfigure{
		\begin{minipage}[t]{0.3\linewidth}
			\centering
			\includegraphics[width=2.3in]{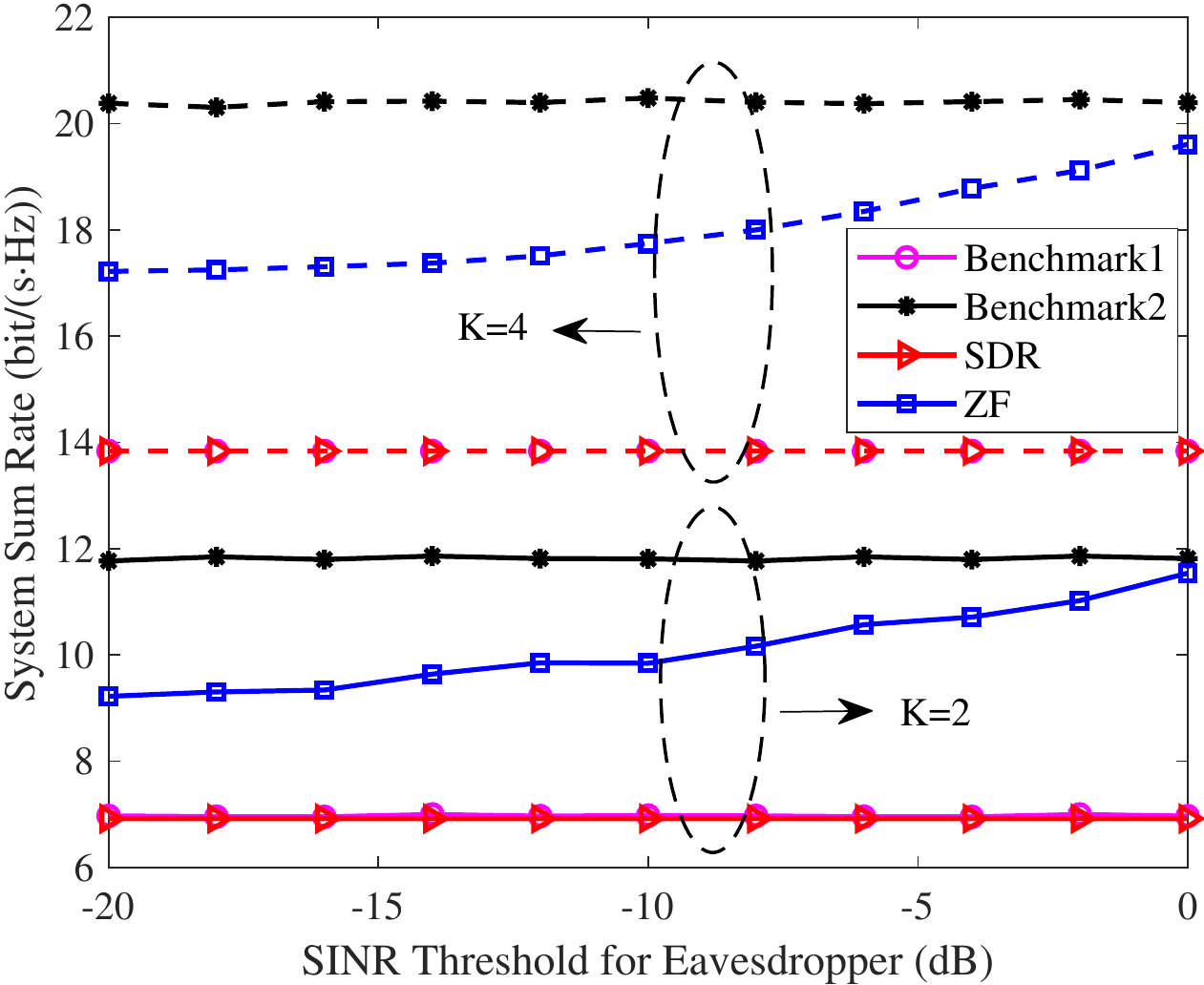}
			\caption{The achievable sum rate versus SINR threshold $\Gamma_e$ for Eves, $\Gamma_c=10$dB.}
			\label{Compare_Eve_SumRate}
	\end{minipage}}
	\hspace{0.01\linewidth} 
	\subfigure{
		\begin{minipage}[t]{0.3\linewidth}
			\centering
			\includegraphics[width=2.3in]{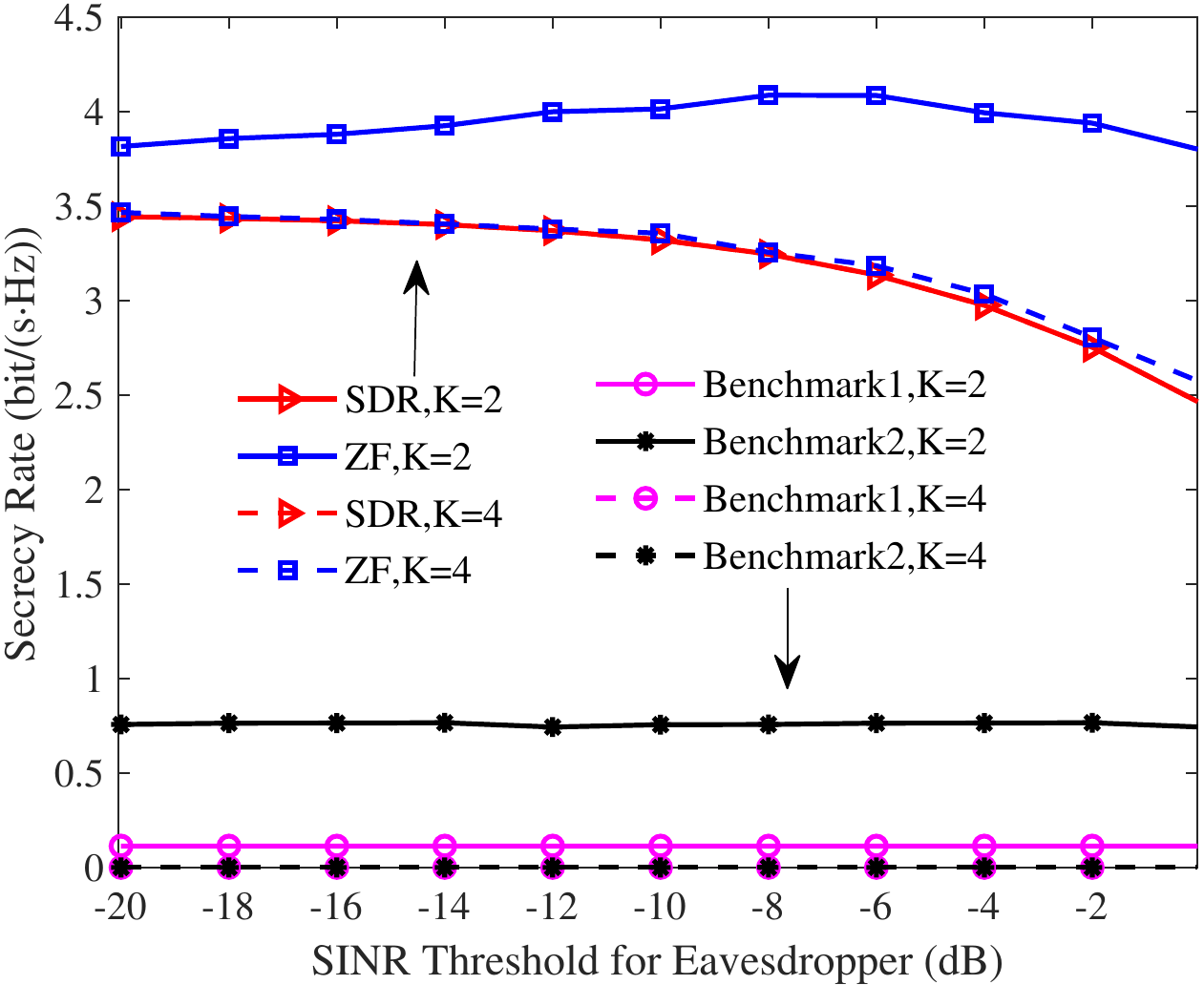}
			\caption{The secrecy rate versus SINR threshold $\Gamma_e$ for Eves, $\Gamma_c=10$dB.}
			\label{Compare_Eve_PLS}
	\end{minipage}}
\end{figure*}

%

In Fig. \ref{Compare_BobTs_SumRate}, we quantify the achievable sum-rate versus the SINR threshold $\Gamma_c$, where the system sum-rate is defined by $\sum_{k=1}^{K}\log_2(1+\gamma_k)$. The SDR and \textit{Benchmark 1} curves are fairly similar and increase linearly with the SINR constraint $\Gamma_c$. This is because the optimal solution should reach the SINR boundary related to the given threshold. Conversely, as seen in the analysis of Section \ref{PAnalysis}, the ZF-based beamformer achieves a higher communication sum rate to the detriment of the radar performance. Meanwhile, the performances of the SDR-based and ZF-based beamformer tend to become similar at high SINR thresholds for both $K=2$ and 4. Furthermore, the curves of the \textit{Benchmark} 2 are slightly higher than those of the proposed ZF algorithms, since there is an additional minimum PLS constraint imposed on the ZF algorithm.

Fig. \ref{Compare_BobTs_PLS} illustrates the system's secrecy rate versus the SINR threshold $\Gamma_c$. Observe that the curves of SDR associated with $K=2$ and $K=4$ are coincident and increase linearly upon increasing $\Gamma_c$. Recall from Section \ref{PAnalysis} that the system's secrecy rate will only reach the value of $\text{log}_2(1+\Gamma_c)-\text{log}_2(1+\Gamma_e)$, if the optimization problem is feasible, regardless of how the other parameters change. Additionally, the ZF-based beamformer associated with $K=2$ achieves a higher secrecy rate than that of the SDR-based algorithm at small values of $\Gamma_c$, since it can reach a higher SINR level than the given threshold. However, the secrecy rate of these two algorithms becomes similar for $K=4$. Actually, supporting more communication users imposes more restrictions on the optimization problem $\mathcal{P}_3$, hence forcing the minimal SINR level to approximate the threshold $\Gamma_c$. Moreover, the proposed PLS-protected beamforming design guarantees a satisfactory PLS level by appropriately choosing the thresholds. By contrast, the benchmark 1 and 2 are not capable of secrecy protection, especially not for numerous legitimate users $K$.


\subsection{System Performance Evaluation vs. the Threshold $\Gamma_e$}\label{SingleT_eve} 

\begin{figure*}[!t]	
	\subfigure{
		\begin{minipage}[t]{0.3\linewidth}
			\centering
			\includegraphics[width=2.3in]{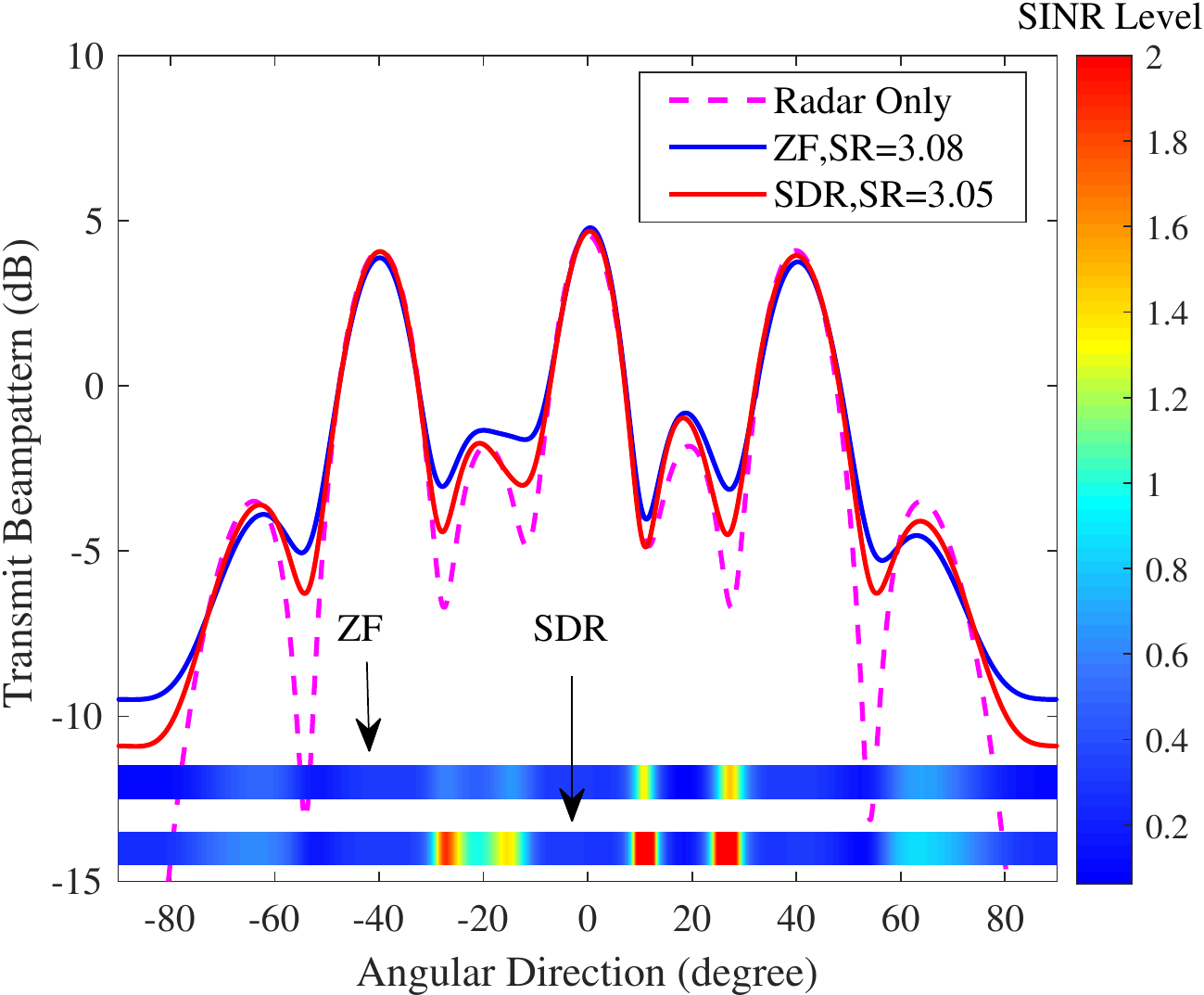}
			\caption{Transmit beampattern for multiple targets with uncertain directions.}
			\label{Compare_MultiplePattern}
	\end{minipage}}
	\hspace{0.01\linewidth}	
	\subfigure{
		\begin{minipage}[t]{0.3\linewidth}
			\centering
			\includegraphics[width=2.3in]{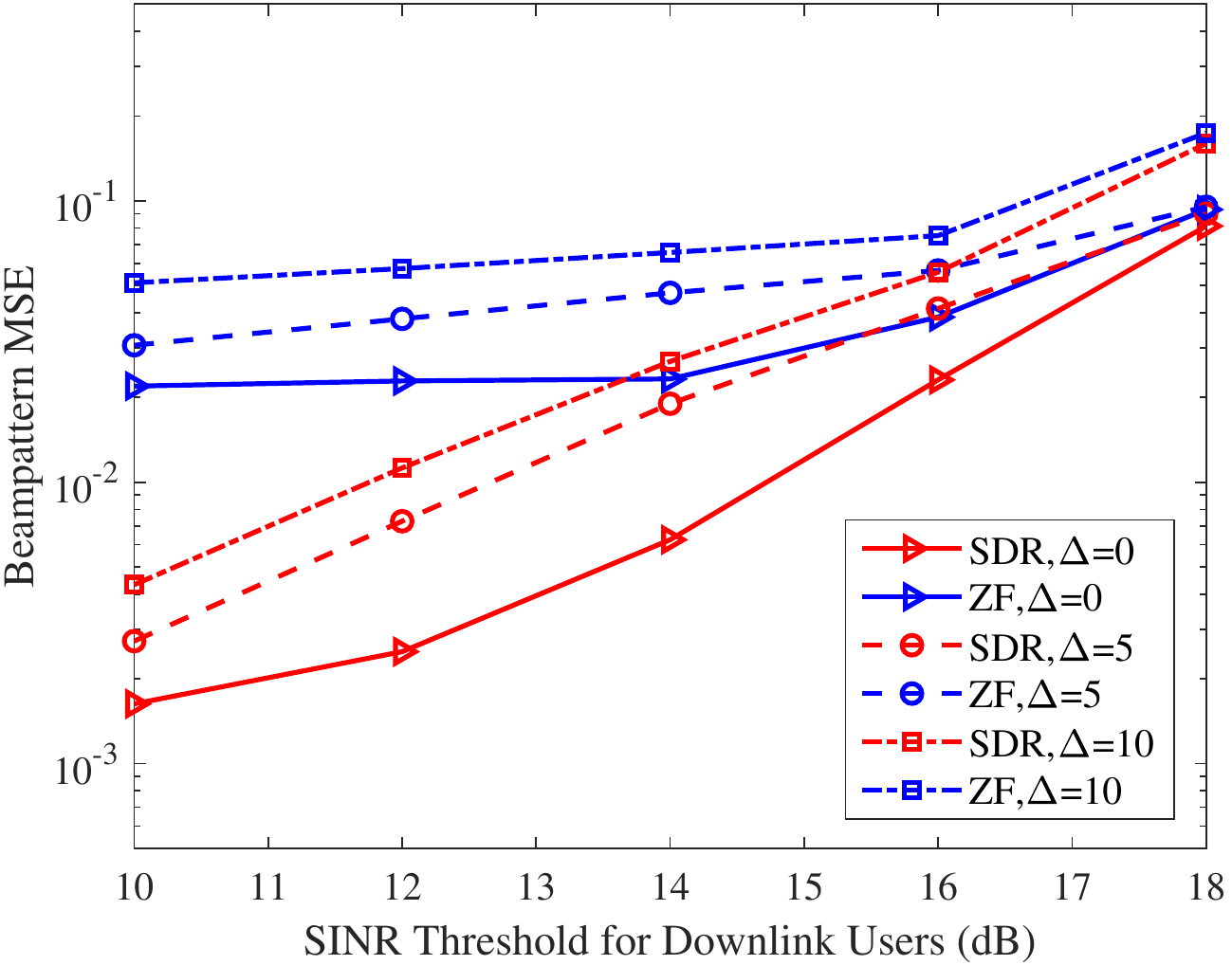}
			\caption{Beampattern MSE comparison with different angular uncertainties of the Eves.}
			\label{Compare_BobTs_MultiMSE}
	\end{minipage}}
	\hspace{0.01\linewidth} 
	\subfigure{
		\begin{minipage}[t]{0.3\linewidth}
			\centering
			\includegraphics[width=2.5in]{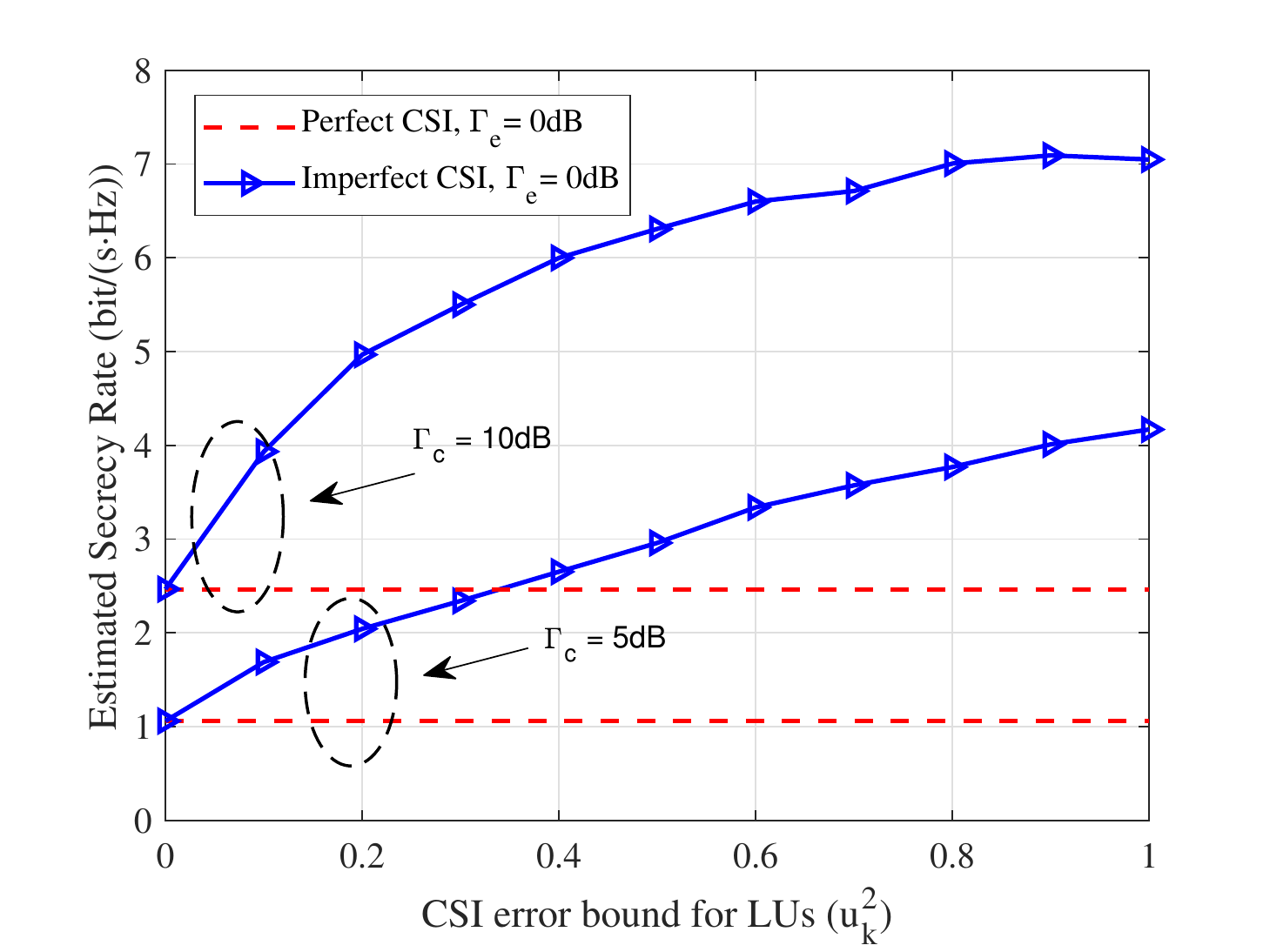}
			\caption{Estimated secrecy rate calculated by the known imperfect CSI versus error bound.}
			\label{CSI_Imperfect}
	\end{minipage}}
	\setcounter{figure}{9}
\end{figure*}

In this subsection, we evaluate the system performance versus the SINR threshold $\Gamma_e$ of the Eves. Accordingly, we set $\Gamma_c=10$dB as a constant, while all other system parameters remain unchanged. The SINR threshold $\Gamma_e$ is varied from $-20$dB to $0$dB with intervals of $2$dB. It should be highlighted that the benchmark curves of \cite{LiuXiang2020} remain constant in all the figures of this subsection. This is because these algorithms do not take the PLS into account, hence the change of threshold $\Gamma_e$ does not affect these performances.

Fig. \ref{Compare_Eve_MSE} shows that the radar beampattern MSE decreases upon increasing $\Gamma_e$ both for the proposed SDR and ZF algorithms. Specifically, we can see that the curves of Fig. \ref{Compare_Eve_MSE} remain near-constant, when $\Gamma_e$ is less than $-12$dB, while decreasing noticeably, when $\Gamma_e$ is higher than $-10$dB. Similar trends may also be observed in Fig. \ref{Compare_Eve_SumRate} and Fig. \ref{Compare_Eve_PLS}, which implies that the performance is not sensitive to the choice of $\Gamma_e$, when $\Gamma_e$ is less than $-12$dB for this set of parameters. Having excessively low $\Gamma_e$ increases the infeasibility probability of the optimization problem considered. 

In Fig. \ref{Compare_Eve_SumRate}, we can see that the system's sum-rate also remains unchanged for the SDR algorithm as a result of the constant threshold $\Gamma_c$ being close to the optimal solution. By contrast, the curves of ZF show an increasing trend in Fig. \ref{Compare_Eve_SumRate} upon increasing $\Gamma_e$, since a higher $\Gamma_e$ implies that less severe restrictions are imposed on the ZF-based beamforming. 

In Fig. \ref{Compare_Eve_PLS}, the SDR and the ZF for $K=4$ reach the boundary of the secrecy rate $\text{log}_2(1+\Gamma_c)-\text{log}_2(1+\Gamma_e)$. Meanwhile, the ZF for $K=2$ attains a higher secrecy rate than its counterparts, since supporting less LUs imposes less restrictions on the beamforming design. Furthermore, we can infer from Fig. \ref{Compare_Eve_SumRate} and Fig. \ref{Compare_Eve_PLS} that although a low $\Gamma_e$ reduces the achievable data rate of the Eve, it also results in a low data rate for the LUs. Therefore, no obvious secrecy rate improvement is attained upon reducing $\Gamma_e$.

\subsection{System Performance Evaluation for imperfect CSI}


First, we evaluate the impact of angular uncertainties of the Eves on the system performance. We set $Q=3$ targets having the directions of $\theta_1=-40^\circ$, $\theta_2=0^\circ$, and $\theta_3=40^\circ$, respectively. Each target has the same direction uncertainty of $\Delta \theta=5^\circ$. The BS detects and tracks these targets, while serving $K=3$ LUs. The SINR thresholds for the LUs and the Eves are set to $\Gamma_c=10$dB and $\Gamma_e=0$dB, respectively. 

Fig. \ref{Compare_MultiplePattern} illustrates the radar transmit beampattern synthesized by the proposed algorithms. The SINR level defined by (\ref{SINREve}) is calculated over the set of $[-90^\circ, 90^\circ]$ angular direction. It can be observed that although the BS forms multi-beams pointing to the directions of the Eves, the SINR levels in each interval covering the Eves are controlled by the threshold $\Gamma_e$. This is because the signal power of radar waveforms is higher than that of the communication symbols, which have to be protected. Moreover, although the beampattern of the ZF algorithm is less beneficial than that of the SDR (higher side-lobe), the average spatial SINR level is lower than that of the SDR algorithm. In Fig. \ref{Compare_BobTs_MultiMSE}, we evaluate the impact of the direction uncertainties on the optimization performance upon varying $\Gamma_c$ from $10$dB to $18$dB. As expected, further constraints are introduced by the uncertainty of the target directions, hence leading to an eroded radar performance.  

Fig. \ref{CSI_Imperfect} shows the estimated secrecy rate calculated by the known imperfect CSI versus the error bound for the scenario of $K=2$. It can be observed that the estimated secrecy rates remain constant and are equal to the secrecy rates in the case of perfect CSI. By contrast, the curves obtained in the case of imperfect CSI exhibit an increasing trend. This is because the worst-case secrecy rate is forced to be larger than a given threshold in our robust beamforming algorithm, while the statistical difference between the worst-case and estimated secrecy rate becomes larger upon increasing the error bound.

\section{Conclusion}\label{Conclusion}
A DFRC multi-user communication system was proposed, while taking the physical layer security into account. The weighted sum of the communication signal and radar waveform was adopted for dual-functional transmission. We demonstrated that the additional radar waveform conveying no confidential information improves the DoF in target detection and simultaneously contaminates the eavesdropping channel. Subsequently, the SDR and the low complexity ZF algorithms were proposed for finding the global optimal solution of the formulated non-convex beamforming design problem. Furthermore, we also designed the robust beamforming for the more practical scenarios of imperfect CSI knowledge. Finally, we evaluated the impact of the parameters on the attainable system performance by numerical simulations, which showed an excellent consistency with the theoretical analysis. Designing PLS systems operating in the face of other types of legitimate and eavesdropping channels as well as hardware impairments is left for our future research. Another promising area of research is the design of Pareto-optimal multi-component systems relying on the full set of optimal operating points in terms of throughput, bit error rate (BER), package loss, latency, etc. 

 \begin{appendices}  
 	\section{The Proof of Proposition 1}\label{A} 
 	By applying the Theorem 1 in \cite{LiuXiang2020}, we only have to prove that the PLS constraint (\ref{cEve}) holds for $\tilde{\textbf{R}},\tilde{\textbf{R}}_1,\cdots,\tilde{\textbf{R}}_K$, if it holds for $\hat{\textbf{R}},\hat{\textbf{R}}_1,\cdots,\hat{\textbf{R}}_K$. First, we show that   
 	\begin{equation}\label{SSS}
 	\textbf{a}^H(\theta)\hat{\textbf{R}}_k\textbf{a}(\theta) \ge \textbf{a}^H(\theta)\tilde{\textbf{R}}_k\textbf{a}(\theta),
 	\end{equation}
 	for arbitrary $\theta$. Upon substituting the expression of $\tilde{\textbf{R}}_k$ into (\ref{optimal}), the right-hand side term of the inequality can be expanded as 
 	\begin{equation}\label{al}
 	\begin{aligned}
 	\textbf{a}^H\tilde{\textbf{R}}_k\textbf{a}&=\textbf{a}^H\tilde{\textbf{w}}_k\tilde{\textbf{w}}_k^H\textbf{a}\\
 	&=(\textbf{h}_k^H\hat{\textbf{R}}_k\textbf{h}_k)^{-1}\textbf{a}^H\hat{\textbf{R}}_k\textbf{h}_k\textbf{h}_k^H\hat{\textbf{R}}_k\textbf{a}\\
 	&=(\textbf{h}_k^H\hat{\textbf{R}}_k\textbf{h}_k)^{-1}|\textbf{a}^H\hat{\textbf{R}}_k\textbf{h}_k|^2.
 	\end{aligned}
 	\end{equation}
 	Additionally, according to the Cauchy-Schwarz inequality, we have
 	\begin{equation}\label{CS}
 	(\textbf{h}_k^H\hat{\textbf{R}}_k\textbf{h}_k)(\textbf{a}^H\hat{\textbf{R}}_k\textbf{a}) \ge |\textbf{a}^H\hat{\textbf{R}}_k\textbf{h}_k|^2.
 	\end{equation}
 	Therefore, it can be readily seen from (\ref{al}) and (\ref{CS}) that (\ref{SSS}) holds. Thus, we can expound as follows
 	\begin{equation}\label{fina}
 	\begin{aligned}
 	\textbf{a}_q^H\tilde{\textbf{R}}\textbf{a}_q+\frac{\sigma_e^2}{|\beta|^2}&\mathop  = \limits^{(a)}\textbf{a}_q^H\hat{\textbf{R}}\textbf{a}_q+\frac{\sigma_e^2}{|\beta|^2}\\
 	&  \ge  (1+\Gamma_e^{-1})\textbf{a}_q^H\sum\limits_{k=1}^K\hat{\textbf{R}}_k\textbf{a}_q\\
 	& \mathop \ge \limits^{(b)} (1+\Gamma_e^{-1})\textbf{a}_q^H\sum\limits_{k=1}^K\tilde{\textbf{R}}_k\textbf{a}_q,
 	\end{aligned}
 	\end{equation}
 	where $(a)$ and $(b)$ follow the first equation in (\ref{optimal}) and the inequality (\ref{SSS}), respectively. Thus, the PLS constraint (\ref{cEve}) holds for $\tilde{\textbf{R}},\tilde{\textbf{R}}_1,\cdots,\tilde{\textbf{R}}_K$, hence completing the proof.
	      
 	\section{The Proof of Proposition 2}\label{B} 
 	The proof is divided into the following three parts: 
 	      	
 	(1) We show that the radar covariance matrix $\hat{\textbf{R}}_\text{rad}$ in (\ref{WrO}) is a positive semidefinite matrix, hence it can be decomposed by either the Cholesky decomposition or by the square root method. Actually, we have
 	\begin{equation} 
 	\begin{aligned}
 	\hat{\textbf{R}}-\textbf{W}_c\textbf{W}_c^H \kern 145pt\\
 	=\hat{\textbf{R}}-\hat{\textbf{R}}_\text{com}+\hat{\textbf{R}}_\text{com}-\textbf{W}_c\textbf{W}_c^H \kern 69pt \\
 	=\hat{\textbf{R}}-\hat{\textbf{R}}_\text{com}+\textbf{L}_c(\textbf{I}-[\textbf{Q}^H]_{[:,1:K]}[\textbf{Q}]_{[1:K,:]})\textbf{L}_c^H.
 	\end{aligned}
 	\end{equation}   
 	Here, $\hat{\textbf{R}}-\hat{\textbf{R}}_\text{com}$ is positive semidefinite due to the constraint (\ref{ZFs}). Since $[\textbf{Q}^H]_{[:,1:K]}$ is the sub-matrix containing the first $K$ columns of unitary matrix, $(\textbf{I}-[\textbf{Q}^H]_{[:,1:K]}[\textbf{Q}]_{[1:K,:]})$ is a positive semidefinite matrix, thereby the last term is also positive semidefinite.
 	
 	(2) We show that the proposed precoding matrices satisfy the ZF constraint (\ref{ZF}). Upon letting $\textbf{F}=\text{diag}(\sqrt{\rho_1},\cdots,\sqrt{\rho_K})$, we have
 	\begin{equation}\label{QR1} 
 	\textbf{H}\textbf{R}_\text{com}\textbf{H}^H=\textbf{H}\textbf{L}_c\textbf{L}_c^H\textbf{H}^H=\textbf{L}_h\textbf{L}_h^H=\textbf{F}\textbf{F}^H.
 	\end{equation}  
 	Note that $\textbf{L}_h\textbf{L}_h^H$ and $\textbf{F}\textbf{F}^H$ are the Cholesky decompositions of the matrix $\text{diag}(\bm{\rho})$, therefore we have $\textbf{L}_h=\textbf{F}$ according to the uniqueness of the Cholesky decomposition of a positive definite matrix. Thus, we have
 	\begin{equation} 
 	\begin{aligned}
 	\textbf{H}\textbf{W}_c&=\textbf{H}\textbf{L}_c[\textbf{Q}^H]_{[:,1:K]}\\
 	&=[\textbf{L}_h,\textbf{0}_{K \times (M-K)}]\textbf{Q}[\textbf{Q}^H]_{[:,1:K]}\\
 	&=\textbf{L}_h=\textbf{F}.
 	\end{aligned}
 	\end{equation} 
 	Moreover, for the radar precoding matrix, we arrive at
 	\begin{equation} \label{000}
 	\begin{aligned}
 	\textbf{H}\textbf{W}_r\textbf{W}_r^H\textbf{H}^H&=\textbf{H}(\hat{\textbf{R}}-\textbf{W}_c\textbf{W}_c^H)\textbf{H}^H\\
 	&=\textbf{F}\textbf{F}^H-\textbf{F}\textbf{F}^H=\textbf{0}.
 	\end{aligned}
 	\end{equation}
 	Thus we can readily obtain $\textbf{H}\textbf{W}_r=\textbf{0}$ from (\ref{000}).  
 	
 	(3) We show that the proposed precoding matrices meet the PLS constraint (\ref{ZFEve}). According to the positive semidefinite property, we can show that
 	\begin{equation} 
 	\textbf{y}^H(\textbf{I}-[\textbf{Q}^H]_{[:,1:K]}[\textbf{Q}]_{[1:K,:]})\textbf{y} \ge 0,
 	\end{equation}
 	for an arbitrary non-zero vector $\textbf{y}$. Upon letting $\textbf{y}=\textbf{L}_c^H\textbf{a}_q$, we have 
 	\begin{equation}\label{1212}
 	\begin{aligned} 
 	\textbf{a}_q^H\textbf{L}_c(\textbf{I}-[\textbf{Q}^H]_{[:,1:K]}[\textbf{Q}]_{[1:K,:]})\textbf{L}_c^H\textbf{a}_q \kern 60pt \\
 	=\textbf{a}_0^H\hat{\textbf{R}}_\text{com}\textbf{a}_q-\textbf{a}_q^H\textbf{W}_c\textbf{W}_c^H\textbf{a}_q \ge 0.
 	\end{aligned}
 	\end{equation}
 	By applying the inequality (\ref{1212}), we can see that
 	\begin{equation}
 	\begin{aligned}
 	\textbf{a}_q^H\hat{\textbf{R}}\textbf{a}_q+\frac{\sigma_e^2}{|\beta|^2} &\mathop  \ge \limits^{(a)} (1+\Gamma_e^{-1})\textbf{a}_q^H\hat{\textbf{R}}_\text{com}\textbf{a}_q\\
 	&  \ge  (1+\Gamma_e^{-1})\textbf{a}_q^H\textbf{W}_c\textbf{W}_c^H\textbf{a}_q, \\
 	\end{aligned}
 	\end{equation}
 	where $(a)$ is valid, because $\hat{\textbf{R}}$ and $\hat{\textbf{R}}_\text{com}$ are the feasible solution of problem $\mathcal{P}_3$ and $\hat{\textbf{R}}$ follows the relationship (\ref{WrO}). Consequently, it can be observed that the precodering matrix constructs also satisfy the PLS constraint, hence completing the proof. 
  
  \end{appendices}

\end{document}